\documentclass[prd,12pt]{revtex4}
\usepackage{amssymb}
\usepackage{amsmath}
\usepackage{graphicx}
\usepackage{color}
\DeclareGraphicsExtensions{.jpg,.jpeg,.pdf,.png,.mps,.eps,.ps}

\newcommand{\beq}{\begin{equation}}
\newcommand{\eeq}{\end{equation}}
\newcommand{\bea}{\begin{eqnarray}}
\newcommand{\ena}{\end{eqnarray}}
\newcommand{\etal}{{\it et al.}}
\newcommand{\ie}{{\it i.e.}}
\newcommand{\eg}{{\it e.g.}}
\newcommand{\lsim}{\mathrel{\mathop{\kern 0pt \rlap
{\raise.2ex\hbox{$<$}}}
\lower.9ex\hbox{\kern-.190em $\sim$}}}
\newcommand{\gsim}{\mathrel{\mathop{\kern 0pt \rlap
{\raise.2ex\hbox{$>$}}}
\lower.9ex\hbox{\kern-.190em $\sim$}}}

\newcommand{\hepth}[1]{{\tt hep-th/#1}}
\newcommand{\hepph}[1]{{\tt hep-ph/#1}}
\newcommand{\physics}[1]{{\tt physics/#1}}

\newcommand{\prep}[3]{#2\ {\it Phys.\ Rep.}\ {\bf #1} #3 }
\newcommand{\plb}[3]{#2\ {\it Phys.\ Lett.\ B}\ {\bf #1} #3 }

\newcommand{\pr}[3]{#2\ {\it Phys.\ Rev.}\ {\bf #1} #3 }
\renewcommand{\prl}[3]{#2\ {\it Phys.\ Rev.\ Lett.} {\bf #1} #3 }
\renewcommand{\prd}[3]{#2\ {\it Phys.\ Rev.\ D}\ {\bf #1} #3 }
\renewcommand{\pra}[3]{#2\ {\it Phys.\ Rev.\ A}\ {\bf #1} #3 }

\newcommand{\jpb}[3]{#2\ {\it J.\ Phys.\ B:\ Atom Mol. Opt. Phys.} {\bf #1} #3 }
\newcommand{\href}[2]{#1}

\definecolor{cyan}{cmyk}{1.,0.,0.,0.5}
\definecolor{magenta}{cmyk}{0.,1.,0.,0.5}
\definecolor{verdatre}{cmyk}{0.5,0.,0.5,0.5}
\definecolor{yellow}{cmyk}{0.,0.,0.2,0.0}
\definecolor{rouge}{cmyk}{0.,0.4,0.6,0.0}
\definecolor{orange}{cmyk}{0.,0.5,0.5,0.}
\definecolor{violet}{rgb}{0.5,0.,0.5}

\begin{document}

\noindent
\title{Reduced Dirac Equation And Lamb Shift As An Off-mass-shell \\
Effect In Quantum Electrodynamics}
 \vskip 1.cm
\author{Ni Guang-jiong $^{\rm a,b}$}
\email{\ pdx01018@pdx.edu}
\affiliation{$^{\rm a}$ Department of Physics, Fudan University, Shanghai, 200433, China\\
$^{\rm b}$ Department of Physics, Portland State University, Portland, OR97207, U. S. A.}

\author{Xu Jianjun}
\email{\ xujj@fudan.edu.cn}
\affiliation{Department of Physics, Fudan University, Shanghai, 200433, China}

\author{Lou Senyue$^{\rm c,d}$}
\email{\ sylou@sjtu.edu.cn}
\affiliation{$^{\rm c}$ Department of Physics, Shanghai Jiao Tong University, Shanghai, 200030, China\\
$^{\rm d}$ Department of Physics, Ningbo University, Ningbo 315211, China}

\vskip 0.5cm
\date{\today}

\vskip 0.5cm
\begin{abstract}
Based on the precision experimental data of energy-level
differences in hydrogenlike atoms, especially the $1S-2S$ transition
of hydrogen and deuterium, the necessity of introducing a reduced
Dirac equation with reduced mass as the substitution of
original electron mass is stressed. Based on new cognition about the essence of special relativity, we
provide a reasonable argument for reduced Dirac equation to have two symmetries, the
invariance under the (newly defined) space-time inversion and that
under the pure space inversion, in a noninertial frame. By using reduced
Dirac equation and within the framework of quantum electrodynamics in covariant
form, the Lamb shift can be evaluated (at one-loop level) as the
radiative correction on a bound electron staying in an
off-mass-shell state--a new approach eliminating the infrared
divergence. Hence the whole calculation, though with limited
accuracy, is simplified, getting rid of all divergences and free
of ambiguity.
{\bf PACC}:\; 0365, 1110G, 1220D
\end{abstract}

\maketitle \vskip 1cm
\section{Introduction}
\label{sec:introduction}

\vskip 0.1cm As is well known, the Dirac equation for electron in
a hydrogenlike atom is usually treated as a one-body equation with
the nucleus being an inert core having infinite mass and exerting
a potential $V(r)=-\frac{Z\alpha}{r}\ \ (\hbar=c=1)$ on the
electron. Then the rigorous solution of energy levels reads[1]:
\bea
 E_{nj}&=&m_ef(n,j)   \label{1-1}\\
 f(n,j)&=&\left[1+\frac{(Z\alpha)^2}{(n-\beta)^2}\right]^{-\frac{1}{2}}\label{1-2}\\
 \beta&=&j+\frac{1}{2}-\sqrt{(j+\frac{1}{2})^2-(Z\alpha)^2}
 \label{1-3}
\ena

where $j$ is the total angular momentum. The expansion of $f(n,j)$ to
the power of $(Z\alpha)^6$ is given as \cite{1}

\beq\begin{array}{l}
f(n,j)=1-\frac{(Z\alpha)^2}{2n^2}-\frac{(Z\alpha)^4}{2n^3}\left(\frac{1}{j+\frac{1}{2}}
-\frac{3}{4n}\right)\\
-\frac{(Z\alpha)^6}{8n^3}\left[\frac{1}{(j+\frac{1}{2})^3}
+\frac{3}{n(j+\frac{1}{2})^2}+\frac{5}{2n^3}-\frac{6}{n^2(j+\frac{1}{2})}\right]+\cdots
\label{1-4}
\end{array}\eeq

Obviously, besides the rest energy of the electron given by the
first term, the second term has exactly the form of Bohr energy
level except that the mass $m_e$ must be replaced by the reduced
mass

\beq \mu=\frac{m_em_N}{m_e+m_N}\equiv\frac{m_em_N}{M}
\label{1-5}
\eeq

with $m_N$ being the mass of the nucleus and $M=m_e+m_N$.

However, as discussed in Refs.\cite{1} and \cite{2}, the concept
of reduced mass in relativistic quantum mechanics (RQM) is
ambiguous to some extent. Beginning from 1950's, a number of
authors have been devoting a great effort at the level of two-body
RQM and that of quantum electrodynamics (QED) to take account of the recoil effect   \cite{3,4,1}, incorporating their
results in a compact form (to order of $\alpha^4$):

\beq
E=M+\mu[f(n,j)-1]-\frac{{\mu}^2}{2M}[f(n,j)-1]^2+\frac{(Z\alpha)^4{\mu}^3}{2n^3m_N^2}
\left[\frac{1}{j+\frac{1}{2}}-\frac{1}{l+\frac{1}{2}}\right](1-{\delta}_{l0})
\label{1-6}
\eeq
A comprehensive review on the theory of hydrogenlike atoms can be found in Ref.\cite{27}.
The aim of this paper is two-fold: First, based on the
experimental data of hydrogen $1S-2S$ transition frequency
\cite{5} and its isotope shift of hydrogen and deuterium \cite{6},
we stress the necessity of the introduction of reduced mass $\mu$
(section II) before we are able to argue the reasonableness of
introducing a "reduced Dirac equation" with $\mu$ as the
substitution of $m_e$ (section III). Second, based on above
conception, we will present a calculation of Lamb Shift (LS) as an
off-mass-shell effect by performing the evaluation of self-energy
diagrams of electron (section IV) and photon (section V) as well
as the vertex function (section VI) at the one-loop level of
QED in covariant form. The new insight
of our calculation is focused on the regularization renormalization
method (RRM). As initiated by J-F Yang \cite{7} and elaborated in
a series of papers (\cite{8:a,8:b,9,24,25,26} and references therein), we can get
rid of all ultra violet divergences in the calculation of quantum
field theory (QFT). Furthermore, in this paper, we will be able to
get rid of the annoying infrared divergence in the vertex function
by treating the electron moving off its mass-shell to certain
extent which is fixed through the evaluation of self-energy
diagram or by the Virial theorem. Based on above improvements, the
one-loop calculation yields values of LS in a simple but
semi-quantitative way (section VII and VIII). Although the
accuracy is limited at one-loop level, we hope our approach could
be served as a new starting point for calculations at high-loop
order to get accurate results at a comparably low labor cost. The
final section IX and Appendix will contain a summary and discussion.
\section{The $1S-2S$ Transition of Atomic Hydrogen and Deuterium}
\label{sec:transition}

\vskip 0.1cm

In the last decade, thanks to remarkable advances in high
resolution laser spectroscopy and optical frequency metrology, the
$1S-2S$ two-photon transition in atomic hydrogen $H$ (or deuterium
$D$) with its natural linewidth of only $1.3 Hz$ had been measured
to a very high precision. In 1997, Udem \etal determined the
$1S-2S$ interval of $H$ being \cite{5}

\beq f^{(H)}(1S-2S)=2466061413187.34(84)\ \ kHz
\label{2-1}
\eeq

Even four years earlier, Schmidt-Kalar \etal ~measured the
isotope-shift of the $1S-2S$ transition of $H$ and $D$ to an
accuracy of $3.7\times 10^{-8}$\cite{6}, giving (as quoted in
\cite{10}):

\beq \Delta f\equiv f^{(D)}(2S-1S)-f^{(H)}(2S-1S)=670994337(22)\ \ kHz
\label{2-2}
\eeq
(In 1998, Huber \etal measured a more accurate data \cite{28}: $670994334.64(15)\ kHz$).
which is of the order of $10^{-4}$ in comparison with Eq.(\ref{2-1}).
As pointed out in Ref.\cite{6}, this $671\ GHz$ isotope-shift can
be ascribed almost entirely to the different masses of proton
($p$) and deuteron ($d$). And the nuclear volume effects become
important because the QED effects cancel considerably in the
isotope shift.

Here, we wish to emphasize that in the first approximation, both
experimental data (\ref{2-1}) and (\ref{2-2}) can be well
accounted for by simply resorting to Eq.(\ref{1-1}) with $m_e$
replaced by the reduced mass

\beq \mu_H=\frac{m_em_p}{m_e+m_p},\ \ \
\mu_D=\frac{m_em_d}{m_e+m_d}
\label{2-3}
\eeq

for $H$ and $D$ respectively.

Indeed, adopting the following updated values \cite{10,11,12,13}

\bea
\alpha&=&(137.03599944)^{-1},\ \ \
{\alpha}^2=0.532513542\times10^{-4}\\
{\alpha}^4&=&0.283570673\times10^{-8},\ \ \
{\alpha}^6=0.151005223\times10^{-12}\\
m_e&=&0.51099906\ \ MeV=1.2355897\times10^{20}\ \ Hz\\
R_{\infty}&=&\frac{1}{2}{\alpha}^2m_e=3.28984124\times10^{15}\ \
Hz\\
\frac{m_p}{m_e}&=&1836.1526665
 \label{data}
 \ena

and denoting
\bea
\frac{m_e}{m_p}&=&b_H=5.446170255\times10^{-4},\ \ \
\frac{1}{1+b_H}=0.999455679\\
\frac{m_e}{m_d}&=&b_D=2.724436319\times10^{-4},\ \ \
\frac{1}{1+b_D}=0.99972763
\label{data2}
\ena

we can calculate the energy difference of $2S$ and $1S$ of $H$
through Eq.(\ref{1-1}) with $m_e$ replaced by $\mu_H$ (the superscript
RDE refers to the reduced Dirac equation)

\bea
\Delta E^{RDE}_{H}(2S-1S)&=&\mu_H[f(2,1/2)-f(1,1/2)]\nonumber \\
&=&\frac{m_e}{1+b_H}(1.996950159\times10^{-5})\nonumber \\
&=&1.2355897\times10^{20}\times0.999455679\times1.996950159\times10^{-5}\nonumber \\
&=&2.466067984\times10^{15}\ \ Hz
\label{2-4}
\ena

which is only a bit larger than the experimental data Eq.(\ref{2-1})
with accuracy $3\times10^{-6}$. However, a more stringent test of
RDE should be the isotope shift of $H$ and $D$. We have

\beq
\frac{1}{1+b_D}-\frac{1}{1+b_H}=(b_H-b_D)-(b^2_H-b^2_D)+(b^3_H-b^3_D)+\cdots=2.719511528\times10^{-4}
\label{2-5}
\eeq

\beq
\Delta
E^{RDE}_{D-H}=(\mu_D-\mu_H)[f(2,1/2)-f(1,1/2)]=6.7101527879\times10^{11}\
\ Hz
\label{2-6}
\eeq

which has only a discrepancy larger than the experimental data,
Eq.(\ref{2-2}) by $20.941\ MHz$ with accuracy $3\times10^{-5}$. Of
course, it is still not satisfied in an analysis of high precision
\cite{6}. Let us resort to the Eq.(6), where the third term does
provide a further modification:

\bea
&-&\frac{1}{2}m_e[\frac{b_D}{(1+b_D)^3}-\frac{b_H}{(1+b_H)^3}]\{[f(2,1/2)-1]^2-[f(1,1/2)-1]^2\}\nonumber\\
&=&\frac{1}{2}m_e[(b_H-b_D)-3(b^2_H-b^2_D)+\cdots](-6.646361554\times10^{-10})=-11.176
\ MHz
\label{2-7}
\ena

which brings the discrepancy between the theory and experimental
down to less than $10\ MHz$.

Although the detail explanation for this discrepancy remains quite
complicated\cite{6}, the above comparison is enough to convince us
that the inevitable appearance of reduced mass in the RDE or
Eq.(6) is by no means a simple fortune. It must have a deep reason
from a theoretical point of view. Notice further that once the
conditions $m_e\ll m_p$ and $m_e\ll m_d$ hold, the difference  of
spin between $p$ and $d$ seems not so important. So in next
section, we will strive to justify the reduced Dirac equation on a
reasonable basis. Of course, it is still an approximate one, but
seems much better than the original Dirac equation when dealing
with hydrogenlike atoms.

\section{Reduced Mass and Reduced Dirac Equation}
\label{sec:decoupling}

Consider a system of two particles with rest masses $m_1$ and
$m_2$. Their coordinates in the center-of-mass (CM) system are ${\mathbf r}_1$ and ${\mathbf r}_2$ respectively,
as shown in Fig.1. If there is a potential $V(r)=V(|{\mathbf
r_1-r_2}|)$ between them, two equations $m_1\ddot{\mathbf
r}_1=-\nabla_rV(r)$ and $m_2\ddot{\mathbf r}_2=\nabla_rV(r)$ will
reduce to one:

\beq
\mu\frac{d^2{\mathbf r}}{dt^2}=-\nabla_rV(r), \ \ \
(\mu=\frac{m_1m_2}{m_1+m_2})
\label{3-1}
\eeq

At first sight, the definition of center-of -mass (CM) in
classical mechanics $m_1r_1=m_2r_2$ becomes doubtful in the theory
of special relativity (SR) because the mass is no longer a
constant. But actually, we can still introduce the coordinate of
CM in the laboratory coordinate system (LCS) (with ${\mathbf
r}'_1$ and ${\mathbf r}'_2$ being the coordinates of $m_1$ and
$m_2$):

\beq
R=\frac{1}{M}(m_1{\mathbf r}'_1+m_2{\mathbf r}'_2)=(X,Y,Z),\
\ \ (M=m_1+m_2)
\label{3-2}
\eeq

and the relative coordinate of $m_1$ and $m_2$ (${\mathbf r}_i={\mathbf r}'_i-{\mathbf R}, \, i=1,2$) :

\beq
{\mathbf r}={\mathbf r}'_1-{\mathbf r}'_2={\mathbf
r}_1-{\mathbf r}_2=(x,y,z)
\label{3-3}
\eeq

Here the motion of CM in the LCS is assumed to be slow and so

\beq
\frac{\partial}{\partial
x'_1}=\frac{m_1}{M}\frac{\partial}{\partial
X}+\frac{\partial}{\partial x},\ \ \ \frac{\partial}{\partial
x'_2}=\frac{m_2}{M}\frac{\partial}{\partial
X}-\frac{\partial}{\partial x}
\label{3-4}
\eeq

Notice that the momentum $\mathbf P$ of CM and the relative
momentum ${\mathbf p}_r$ becomes operator in quantum mechanics
(QM) without explicit dependence on mass:

\beq
{\mathbf P}=-i\hbar\nabla_{\mathbf R},\ \ \ {\mathbf
p}_r=-i\hbar\nabla_{\mathbf r}
\label{3-5}
\eeq

Thus the momenta of $m_1$ and $m_2$ in laboratory coordinate
system (LCS) read:

\beq
{\mathbf p}'_1=-i\hbar\nabla_{{\mathbf
r}'_1}=\frac{m_1}{M}{\mathbf P}+{\mathbf p}_r,\ \ \ {\mathbf
p}'_2=-i\hbar\nabla_{{\mathbf r}'_2}=\frac{m_2}{M}{\mathbf
P}-{\mathbf p}_r
\label{3-6}
\eeq

Since the center-of-mass coordinate system (CMCS) is also an inertial frame which can be transformed
from the LCS via a linear Lorentz transformation, it is defined by
the condition that $\mathbf P=0$ in CMCS. In other words, CMCS is
defined by the condition ${\mathbf p}_1+{\mathbf p}_2=0$, or from
Eq.(\ref{3-6}):

\beq
{\mathbf p}'_1=-i\hbar\nabla_{{\mathbf r}'_1}={\mathbf p}_r,\
\ \ {\mathbf p}'_2=-i\hbar\nabla_{{\mathbf r}'_2}=-{\mathbf p}_r
\label{3-7}
\eeq

Evidently, the above definition of CMCS remains valid in the realm
of relativistic QM (RQM) even the exact meaning of CM seems
obscure to some extent due to the conjugation relation of $a$ particle's position and its momentum, see Fig.1.

Now, from Eq.(\ref{3-7}), it is natural to replace $\mathbf p_1$ and
$\mathbf p_2$ by ${\mathbf p}_r$, reducing the two-particle
degrees of freedom to one. In the meantime, the origin of CMCS is
discarded, it is substituted  by the position of $m_2$ (${\mathbf
r}={\mathbf r}_1-{\mathbf r}_2$). We will call the system
associated with $\mathbf r$ the relative motion coordinate system
(RMCS), which should be viewed as a deformation of CMCS. The
transformation from CMCS to RMCS is by no means a linear one.
Rather, the origin of RMCS ($m_2$) is moving non-uniformly in the CMCS.
Therefore, while rest masses $m_1$ and $m_2$ remain the same in
both LCS and CMCS, they reduce to one mass
$\mu=\frac{m_1m_2}{m_1+m_2}$ for $m_1$ in RMCS (or for $m_2$ if
$m_1$ is chosen as the origin of RMCS).

Let us express the total energy $E=E_1+E_2$ in CMCS in terms of
$p_r$ and reduced mass $\mu$ ($\mu=\frac{m_1m_2}{M},\
M=m_1+m_2$), where

\beq
E_1=\sqrt{m_1^2+p_1^2}=\sqrt{m_1^2+p_r^2},\ \ \
E_2=\sqrt{m_2^2+p_2^2}=\sqrt{m_2^2+p_r^2}
\label{3-8}
\eeq

Treating all $p_1,p_2$ and $p_r$ being $c$-numbers, we have

\beq
E^2=(E_1+E_2)^2=M^2+\frac{M}{\mu}p_r^2+\frac{1}{4\mu^2}p_r^4(4-\frac{M}{\mu})+\cdots
\label{3-9}
\eeq
where the expansion in $p_r$ is kept to the order
of $p_r^4$. Two extreme cases will be considered separately:

{\bf A}. $m_2\gg m_1,\ \  \mu\lesssim m_1, \ \ M\gg\mu$: \bea
E^2&=&M^2\left[1+\frac{1}{\mu
M}p_r^2-\frac{1}{4M\mu^3}p_r^4(1-\frac{4\mu}{M})+\cdots\right]\nonumber\\
E&=&M\left[1+\frac{1}{2\mu
M}p_r^2-\frac{1}{8M\mu^3}p_r^4(1-\frac{3\mu}{M})+\cdots\right]\nonumber\\
&=&M+\frac{1}{2\mu}p_r^2-\frac{1}{8\mu^3}p_r^4+\cdots\nonumber\\
&\simeq &M-\mu+\sqrt{\mu^2+p_r^2}\simeq m_2+(m_1-\mu)+\sqrt{\mu^2+p_r^2}
\label{3-10}
\ena

\beq
E'\equiv E-m_2=(m_1-\mu)+\sqrt{\mu^2+p_r^2}
\label{3-11}
\eeq

{\bf B}. $m_1=m_2=m,\ \ \mu=\frac{m}{2},\ \  M=2m=4\mu$
Then to
the accuracy of $p_r^4$, we have :

\bea
E^2&=&M^2+\frac{M}{\mu}p_r^2=4m^2+4p_r^2\nonumber\\
E&=&2m+\frac{1}{2\mu}p_r^2-\frac{1}{32\mu^3}p_r^4+\cdots\nonumber\\
E'&\equiv &E-M=\frac{1}{2\mu}p_r^2-\frac{1}{32\mu^3}p_r^4\simeq
\frac{1}{2\mu}p_r^2,\ \ \ (\mbox{if}\ \ p_r^2\ll \mu^2)
\label{3-12}
\ena

It is interesting to see that after introducing $\mu$ and $p_r$,
the energy $E'$ in RMCS looks  quite "relativistic" in the case A
whereas it looks rather "non-relativistic" in the case B even both
of them are derived from the relativistic expressions, Eq.(\ref{3-8}),
approximately.

Since the RMCS is not an inertial system, the original mass of
$m_1$ in CM changes abruptly to $\mu$ as shown in Eq.(\ref{3-11}).
How can we derive the reduced Dirac equation (RDE) in RMCS?
Fortunately, we already found a basic symmetry, the space-time
inversion symmetry, which not only serves as the essence of
special relativity (SR), but also goes beyond it to derive the original Dirac equation and the
tachyon theory for neutrinos \cite{14,15,16,17}. Based on this symmetry,
we are going to derive the equation in RQM for case either A or B
respectively.

Let us consider case B ($m_1\simeq m_2$) first. The motivation is
stemming from the success of using the Schr\"{o}dinger equation to
heavy-quarkoniums like $c\bar{c}$ and $b\bar{b}$ in particle
physics (\cite{18}, see also \cite{15} \ \S 9.5 D). Ignoring the
spin of both $m_1$ and $m_2$, we assume the coupling equations
in laboratory system for the two-particle system as:

\beq
\left\{
\begin{array}{ll}
i\hbar\frac{\partial\varphi}{\partial
t}&=(m_1+m_2)c^2\varphi+V(|{\mathbf
r'_1-\mathbf r'_2}|)(\varphi+\chi)-(\frac{\hbar^2}{2m_1}\nabla^2_{\mathbf
r'_1}+\frac{\hbar^2}{2m_2}\nabla^2_{\mathbf r'_2})(\varphi+\chi)\\
i\hbar\frac{\partial\chi}{\partial
t}&=-(m_1+m_2)c^2\chi-V(|{\mathbf
r'_1-\mathbf r'_2}|)(\varphi+\chi)+(\frac{\hbar^2}{2m_1}\nabla^2_{\mathbf
r'_1}+\frac{\hbar^2}{2m_2}\nabla^2_{\mathbf r'_2})(\varphi+\chi)
\end{array}
\right.
\label{3-13}
\eeq

where $\varphi=\varphi({\mathbf r'_1,\mathbf r'_2},t)$ and
$\chi=\chi({\mathbf r'_1,\mathbf r'_2},t)$ are hidden "particle" and
"antiparticle" fields of the two-particle system (From now on, the ${\mathbf r}'_i(i=1,2)$ is the flowing coordinate of "fields"
in QM, \ie, that of "fictitious point particles". See Fig.1).
Eq.(\ref{3-13}) remains invariant under the (newly defined) space-time inversion
(${\mathbf r'_1\rightarrow-\mathbf r'_1,
\mathbf r'_2\rightarrow-\mathbf r'_2},t\rightarrow-t$):

\beq \left\{
\begin{array}{ll}
\varphi({-\mathbf r'_1,-\mathbf r'_2},-t)&\longrightarrow\chi({\mathbf
r'_1,\mathbf r'_2},t)\\
\chi({-\mathbf r'_1,-\mathbf r'_2},-t)&\longrightarrow\varphi({\mathbf
r'_1,\mathbf r'_2},t)
\end{array}\right.
\label{3-14}
\eeq

\beq
V({-\mathbf r'_1,-\mathbf r'_2},-t)\longrightarrow V({\mathbf r'_1,\mathbf r'_2},t)
\label{3-15}
\eeq

Note that, however, the time $t$ is not contained in $V$ explicitly. Eq.(\ref{3-15}) merely means that both $m_1$ and $m_2\ (m_1\approx m_2)$
transform into their antiparticles under the space-time inversion.
Actually, the hidden antiparticle field $\chi$ enhances in nearly
equal strength in $m_1$ and $m_2$ when fictitious particles' velocities
increase with the enhancement of attractive potential $V(r)$.

After introducing the CM coordinate ${\mathbf
R}=\frac{1}{M}(m_1{\mathbf r'}_1+m_2{\mathbf r'}_2)$, ($M=m_1+m_2$)
and relative coordinate $\mathbf r=\mathbf r'_1-\mathbf r'_2$, and setting

\beq
\varphi=\Phi+i\frac{\hbar}{Mc^2}\dot{\Phi},\ \ \ \
\chi=\Phi-i\frac{\hbar}{Mc^2}\dot{\Phi}
\label{3-16}
\eeq

we find ($\mu=\frac{m_1m_2}{M}$)

\beq
\ddot{\Phi}-c^2\nabla^2_R\Phi-c^2\frac{M}{\mu}\nabla^2_r\Phi+\frac{1}{\hbar^2}(M^2c^4+2VMc^2)\Phi=0
\label{3-17} \eeq

Its stationary solution reads

\beq
\Phi({\mathbf R,\mathbf r},t)=\psi(\mathbf
r)\exp\left[\frac{i}{\hbar}({\mathbf P\cdot \mathbf R}-Et)\right]
\label{3-18}
\eeq

where $E$ is the total energy of the system while $\mathbf P$ the
momentum of CM. The reduced "one-body" equation for $\psi(\mathbf r)$ turns out to
be: \footnotemark[1]
\footnotetext[1]{With Eq.(\ref{3-15}), Eq.(\ref{3-17}) is invariant under the space-time inversion (${\bf r}\to -{\bf r},
t\to -t$). Equivalently, under the mass inversion ($m_1\to -m_1,m_2\to -m_2$), Eq.(\ref{3-17}) and Eq.(\ref{3-19}) remain
invariant in the sense that not only $\mu\to -\mu, M\to -M$, but also $V({\bf r})\to -V({\bf r}), \varepsilon\to -\varepsilon$.
Notice that, however, the simultaneous inversion of $m_1$ and $m_2$ implies $m_1\simeq m_2$, so both particles change under their
mutual interaction $V({\bf r})$ simultaneously. Here $V$, being the "internal potential energy" of two-body system, was called
as a "scalar potential". We see that either the invariance under the space-time inversion or that under the mass inversion
is capable of showing the particle-antiparticle symmetry (\ie, relativistic nature) of a system essentially. }

\beq
\left\{
\begin{array}{ll}
&\left[-\dfrac{\hbar^2}{2\mu}\nabla^2_{\mathbf r}+V(\mathbf r)\right]\psi(\mathbf
r)=\varepsilon\psi(\mathbf r)\\[4mm]
&\varepsilon=\dfrac{1}{2Mc^2}(E^2-M^2c^4-{\mathbf P}^2c^2)
\end{array}
\right.
\label{3-19}
\eeq

We set $\mathbf P=0$ (\ie  turn to CMCS) and denote the binding
energy $B=Mc^2-E$, yielding:

\beq
B=Mc^2\left[1-(1+\frac{2\varepsilon}{Mc^2})^{1/2}\right]=-\varepsilon+\frac{1}{2}\frac{\varepsilon^2}{Mc^2}-\cdots
\label{3-20}
\eeq

Notice that although Eq.(\ref{3-19}) looks like a
"non-relativistic" stationary Schr\"{o}dinger equation, it is
essentially relativistic. This can be seen from its remarkable
property that the eigenvalue $\varepsilon$ has a lower bound
$-\frac{1}{2}Mc^2$, corresponding to $E_{\mbox{min}}=0\
(B_{\mbox{max}}=M)$!

An example is: consider "positronium" composed of $e^+$ and $e^-$ with charge $Ze$ and $-Ze$ respectively. Once when the "fictitious
charge number" $Z$ increases from $1$ to $Z_{max}=(\frac{4}{\alpha^2})^{1/4}=16.555$, the whole bound system would have lowest ground
energy $E_{min}=0$!
So Eq.(\ref{3-19}) is really a relativistic QM equation capable of
giving a nonperturbative solution under the strong coupling.

Eq.(\ref{3-19}) provides a justification (realization) of
conjecture Eq.(\ref{3-12}) relevant to case B ($m_1\simeq m_2$) where
the spin of both particles is merely of second importance.

Now let us turn to case A where $m_2\gg m_1$, taking the spin of
$m_1$ into account but ignoring that of $m_2$ as before. Based on
the experience in case B, also because of great difficulty to
derive the equation starting from the laboratory system for this
case A, we directly introduce the reduced Dirac equation (RDE) in
the RMCS as a pair of coupled equations of two-component spinors
$\varphi({\mathbf r},t)$ and $\chi({\mathbf r},t)$, ($c=\hbar=1$)

\beq
\left\{
\begin{array}{ll}
i\dot{\varphi}&=i{\mathbf \sigma}_1\cdot\nabla_{\mathbf r}\chi+\mu\varphi+V({\mathbf r})\varphi\\
i\dot{\chi}&=i{\mathbf \sigma}_1\cdot\nabla_{\mathbf r}\varphi-\mu\chi+V({\mathbf r})\chi
\end{array}
\right.
\label{3-21}
\eeq
with $\mu$ replacing $m_1$. Here $\mathbf \sigma_1$ are Pauli matrices acting on the spin
space of particle $m_1$. Eq.(\ref{3-21}) is invariant under the
space-time inversion $({\mathbf r\rightarrow-\mathbf r},t\rightarrow-t), \
\varphi(-{\mathbf r},-t)\rightarrow \chi({\mathbf r},t),\ \chi(-{\mathbf r},-t)\rightarrow \varphi({\mathbf r},t)$
whereas we assume

\beq
V(-{\mathbf r},-t)\longrightarrow-V({\mathbf r},t)
\label{3-22}
\eeq
here in contrast to Eq.(\ref{3-15}) for the case B. \footnotemark[2]
\footnotetext[2]{For a hydrogenlike atom, $V(r)=-\frac{Ze^2}{r}$ does not contain time $t$ explicitly.
Eq.(\ref{3-22}) merely means that under the space-time inversion, the electron transforms into a position whereas the nucleus
remains unchanged. See point (a) of section IX. Previously, the $V$ in Eq.(\ref{3-22}) was called as a "vector potential",
meaning the "potential energy" of the electron in an "external field" of nucleus. Note that, formally, Eq.(\ref{3-21}) remains
invariant under a mass inversion as $\mu\to -\mu, \phi\to\chi, \chi\to\phi$ ($V({\bf r})$ remains unchanged) in a noninertial
frame $RMCS$. Actually, since $m_1=m_e\to -m_e$, but $m_2=m_N\to m_N, \mu\to -\mu(1+\frac{2m_e}{M})$. So Eq.(\ref{3-21})
has an inaccuracy up to $\frac{2m_e}{M}$ ($<1.1\times 10^{-3}$ for $H$).}

The reasons are as follows: (a) Eq.(\ref{3-21}) should degenerate into the
original Dirac equation when $m_2\rightarrow\infty$,
$\mu=\frac{m_1m_2}{m_1+m_2}\rightarrow m_1$. (b) Since now $m_2\gg
m_1$ (but $m_2\neq\infty$), $m_1$ is moving much faster than $m_2$
in the CMCS. Hence the antiparticle field $\chi$ enhances much
appreciably in $m_1$ than that in $m_2$, a situation totally
different from that in the case B where $m_1\approx m_2$. (c) If
instead of Eq.(\ref{3-22}), we still assume $V(-{\mathbf
r},-t)\longrightarrow V({\mathbf r},t)$ like Eq.(\ref{3-15}) and
change the sign before $V(\mathbf r)$ in the second equation of
Eq.(\ref{3-21}) to keep its invariance under the space-time
inversion, then we would get an equation which would lead to a
reversed fine-structure of atom (\eg, the $P_{1/2}$ state would
lie above the $P_{3/2}$ state), a wrong prediction obviously
excluded by experiments.

However, one kind of invariance is not enough to fix an equation. Indeed,
the beauty of Dirac equation or RDE is hidden in two symmetries:
besides the symmetry of space-time inversion, it has another
left-right (parity) symmetry. To see it, we define

\beq
\xi=\frac{1}{\sqrt{2}}(\varphi+\chi),\ \ \
\eta=\frac{1}{\sqrt{2}}(\varphi-\chi)
\label{3-23}
\eeq

and recast Eq.(\ref{3-21}) into:

\beq
\left\{
\begin{array}{ll}
i\dot{\xi}&=i{\mathbf \sigma}_1\cdot\nabla_{\mathbf r}\xi+\mu\eta+V(\mathbf r)\xi\\
i\dot{\eta}&=-i{\mathbf \sigma}_1\cdot\nabla_{\mathbf r}\eta+\mu\xi+V(\mathbf r)\eta
\end{array}
\right.
\label{3-24}
\eeq

which is invariant under a pure space inversion (${\mathbf
r\rightarrow-\mathbf r},t\rightarrow t$) if assuming

\beq
\xi({-\mathbf r},t)\rightarrow\eta({\mathbf r},t),\ \
\eta({-\mathbf r},t)\rightarrow\xi({\mathbf r},t),\ \ V(-{\mathbf
r})\rightarrow V({\mathbf r})=V(r)
\label{3-25}
\eeq

The parity invariance of Dirac equation or RDE has a far-reaching
consequence that the Dirac particle is always a subluminal
one. By contrast, once the parity is violated to maximum, a
superluminal particle (tachyon) will emerge. Interestingly enough,
any theory capable of treating particle and antiparticle on an equal
footing must respect to the common basic symmetry---the
invariance of space-time inversion. The new insight of this
section is this symmetry can be applied even in a noninertial
frame---the RMCS. Of course, the validity of RDE can only be
verified by experiments as discussed in section II, although it is
still an approximate description of nature like any other theory
in physics. For further discussion, see section IX.

\section{Self-Energy Correction of a Bound Electron in Atom}
\label{sec:selfenergy}

In our understanding, one important reason why the calculations of
QED for electron in a hydrogenlike atom is so complicated lies in
the fact that while calculations are performed in the CMCS, the
center of potential (the nucleus with mass $m_2=m_N$) undergoes a
complex motion. So the recoil effect interwinds with the high-loop
correction of QED, as discussed in many chapters of the books
\cite{1} and \cite{2}. We will try to find an alternative approach
by adopting the RDE and doing calculation in the RMCS. Let us
begin with the Feynman diagram integral (FDI) of electron
self-energy at one-loop level, adopting the Bjorken-Drell metric
and rationalized Gaussian units with electron charge $-e(e>0)$,
see Fig.2(a) (\cite{8:a}).

\beq
-i\Sigma(p)=(ie)^2\int\frac{d^4k}{(2\pi)^4}\frac{g_{\mu\nu}}{ik^2}
\gamma^{\mu}\frac{i}{\not\!\!p-\not\!\!k-\mu}\gamma^{\nu}
\label{4-1}
\eeq

Here a free electron with reduced mass $\mu$ is moving at a
four-dimensional momentum $p$, whose spatial component is just the
relative momentum $\mathbf p_r$ discussed in the previous section,
$k$ is the momentum of virtual photon. As usual, a Feynman parameter $x$ will bring
Eq.(\ref{4-1}) into

\beq
-i\Sigma(p)=-e^2\int\frac{d^4k}{(2\pi)^4}\frac{N}{D}
\label{4-2}
\eeq

\beq
\frac{1}{D}=\int^1_0\frac{dx}{[k^2-2p\cdot
kx+(p^2-\mu^2)x]^2},\ \ \ N=-2(\not\!\!p-\not\!\!k)+4\mu
\label{4-3}
\eeq

($\not\!\!p=p^\mu\gamma_\mu,\ p\cdot k=p^\mu k_\mu$). A shift in
momentum integration:$k\rightarrow K=k-xp$ recast Eq.(\ref{4-2})
into

\beq
-i\Sigma(p)=-e^2\int^1_0dx[-2(1-x)\not\!\!p+4\mu]I
\label{4-4}
\eeq

with a logarithmically divergent integral (in Minkowski momentum
space):

\beq I=\int\frac{d^4K}{(2\pi)^4}\frac{1}{[K^2-M^2]^2},\ \ \
M^2=p^2x^2+(\mu^2-p^2)x \label{4-5} \eeq

Our new regularization-renormalization method (RRM) is based on a
cognition that the virtual process in the self-energy diagram does provide a radiative correction to the
electron mass but only when the electron is off the mass shell,
\ie, $p^2\not=\mu^2$. When it is on the mass shell, $p^2=\mu^2$,
the appearance of a divergent integral like $I$ in Eq.(\ref{4-5}) is
essentially a warning on the fact that to calculate the mass of
electron is beyond the ability of perturbative QED.

Let us consider the converse: if $\Sigma(p)$ does modify the
electron mass $\mu$ to some extent, it must comes from the
divergent integral $I$. However, the latter is a dimensionless
number, we can change the unit of $M$ (and $k$) at our disposal
without any change in the value of $I$. So any real change of
$\mu$ (on the mass shell) is incredible. The deeper reason lies in
a "principle of relativity" in epistemology: everything is moving
and becomes recognizable only in relationship with other things. What we can understand is either no
mass scale or two mass scales, but never one mass scale.
For instance, in the famous Gross-Neveu model \cite{19}, a massive
fermion is created only in accompanying with the change (phase
transition) of its environment (vacuum) which provides another
mass scale (a standard weight). Another example is just the change
of electron mass from $m_e$ to $\mu_H$ in a hydrogen atom due to
the coexistence of atom nucleus---the proton, this change is also
a nonperturbative effect.

Therefore, we expected too much in the past. There is no way to evaluate Eq.(\ref{4-5}) unambiguously
or pick out some finite and fixed modification on the mass $\mu$.
What we can do is to separate the valuable information carried by
Eq.(\ref{4-5}) from an arbitrary constant which will be introduced by a simple trick and then
fixed by the experimental data of $\mu$. We will see the
information telling us exactly how the value of $I$ changes when
the electron is moving off the mass shell.

To handle Eq.(\ref{4-5}), we perform a differentiation with respect
to the mass-square parameter $M^2$, then the integration with
respect to $K$ becomes convergent, yielding:

\beq
\frac{\partial I}{\partial
M^2}=\frac{-i}{(4\pi)^2}\frac{1}{M^2}
\label{4-6}
\eeq

which tells us that while the exact value of $I$ remains obscure,
its change linked with $M^2$ has a definite meaning. So we
reintegrate Eq.(\ref{4-6}) with respect to $M^2$ and arrive at

\beq I=\frac{-i}{(4\pi)^2}(\ln
M^2+C_1)=\frac{-i}{(4\pi)^2}\ln\frac{M^2}{\mu_2^2} \label{4-7}
\eeq

where an arbitrary constant $C_1=-\ln\mu_2^2$ is introduced
($\mu_2$ should not be confused with the reduced mass $\mu$).
Further integration with respect to Feynman parameter $x$ leads to

\beq
\begin{array}{ll}
 \Sigma(p)&=A+B\not\!\!p\\
A&=\frac{\alpha}{\pi}\mu\left[2-2\ln\frac{\mu}{\mu_2}+\frac{(\mu^2-p^2)}{p^2}\ln\frac{(\mu^2-p^2)}{\mu^2}\right]\\
B&=\frac{\alpha}{4\pi}\left[2\ln\frac{\mu}{\mu_2}-3-\frac{(\mu^2-p^2)}{p^2}
\left[1+\frac{(\mu^2+p^2)}{p^2}\ln\frac{(\mu^2-p^2)}{\mu^2}\right]\right]
\end{array}
\label{4-8}
\eeq

Using the chain approximation, we can derive the modification of
electron propagator as

\beq
\frac{i}{\not\!\!p-\mu}\rightarrow\frac{i}{\not\!\!p-\mu}\frac{1}{1-\frac{\Sigma(p)}{\not\!\!p-\mu}}
=\frac{iZ_2}{\not\!\!p-\mu_R}
\label{4-9}
\eeq

where

\beq
Z_2=\frac{1}{1-B}
\label{4-10}
\eeq

is the renormalization factor for wave function of electron and

\beq
\mu_R=\frac{\mu+A}{1-B}
\label{4-11}
\eeq

is the renormalized mass of $\mu$. The increment of mass reads

\beq
\delta\mu=\mu_R-\mu=\frac{A+\mu B}{1-B}
\label{4-12}
\eeq

For a free electron (in the atom), the mass-shell condition
$p^2=\mu^2$ should lead to

\beq
\delta\mu|_{p^2=\mu^2}=\frac{\alpha\mu}{4\pi}(5-6\ln\frac{\mu}{\mu_2})=0
\label{4-13}
\eeq

as discussed above\footnotemark[3]. So we must set $\mu_2=\mu e^{-5/6}$ which in
turn fixes
\footnotetext[3]{We will keep the same mass symbol $\mu$ through out high-loop calculations of QED and reconfirm (renormalize)
it at every step by experiment. Just like one has to reconfirm his plane ticket before his departure from the airport,
he must use the same name through out his entire journey \cite{8:b}.}
\beq
Z_2|_{p^2=\mu^2}=\frac{1}{1+\frac{\alpha}{3\pi}}\approx
1-\frac{\alpha}{3\pi}
\label{4-14}
\eeq

However, the above evaluation further provides us with important
knowledge of $\delta\mu$ when electron is moving off the
mass-shell. Consider the similar diagram in Fig.(2b), we can
set on an average meaning that

\beq
p^2=\mu^2(1-\zeta)
\label{4-15}
\eeq

with $\zeta>0$, which implies from Eq.(\ref{4-12}) with Eq.(\ref{4-8})
that \cite{20}:

\beq
\delta\mu=\frac{\alpha\mu}{4\pi}\frac{(-\zeta+2\zeta\ln\zeta)}{1+\alpha/3\pi}
\label{4-16}
\eeq

where some terms of the order of $\zeta^2$ or $\zeta^2\ln\zeta$ are
neglected since $\zeta\ll1$. Eq.(\ref{4-16}) establishes the
correspondence between the mass modification $\delta\mu$ and the
parameter $\zeta$ describing the off-mass-shell extent of electron
in the bound state. For a hydrogenlike atom, we may ascribe
$\delta\mu$ to the (minus) binding energy of electron in the Bohr
theory:

\beq
\delta\mu=\varepsilon_n=-\frac{Z^2\alpha^2}{2n^2}\mu
\label{4-17}
\eeq

Then Eq.(\ref{4-16}) gives the value of $\zeta$ for fixed values $Z$
and $n$. We will see from the vertex function that these values of
$\zeta$ are crucial to the calculation of Lamb shift (sections
VII and VIII).

\section{Photon Self-energy}

As discussed in various text books \cite{21,22,23}, we encounter
the FDI of vacuum polarization Fig.2(c) as \cite{8:a}:

\beq \Pi_{\mu\nu}(q)=-(-ie)^2Tr\int\frac{d^4\bar{p}}{(2\pi)^4}
\gamma_{\mu}\frac{i}{\not\!\!\bar{p}-m}\gamma_{\nu}\frac{i}{\not\!\!\bar{k}+\not\!\!q-m}
\label{5-1}
\eeq

Here $q$ is the momentum transfer along the photon line and $m$ the mass of electron.
Introducing the Feynman parameter $x$ as in previous section and performing a shift in momentum integration:
$\bar{p}\rightarrow K=\bar{p}+xq$, we get:

\beq \Pi_{\mu\nu}(q)=-4e^2\int_0^1dx(I_1+I_2)
\label{5-2}
\eeq

where

\beq
I_1=\int\frac{d^4K}{(2\pi)^4}\frac{2K_{\mu}K_{\nu}-g_{\mu\nu}K^2}{(K^2-M^2)^2}
\label{5-3}
\eeq

with

\beq
M^2=m^2+q^2(x^2-x)
\label{5-4}
\eeq

is quadratically divergent while

\beq
I_2=\int\frac{d^4K}{(2\pi)^4}\frac{(x^2-x)(2q_{\mu}q_{\nu}-g_{\mu\nu}q^2)+m^2g_{\mu\nu}}{(K^2-M^2)^2}
\label{5-5}
\eeq

is only logarithmically divergent like that in Eq.(\ref{4-5}). An
elegant way to handle $I_1$, Eq.(\ref{5-3}), is modifying $M^2$ to

\beq
M^2(\sigma)=m^2+q^2(x^2-x)+\sigma
\label{5-6}
\eeq

and differentiating $I_1$ with respect to $\sigma$.
After integration with respect to $K$, we reintegrate it with
respect to $\sigma$ twice, arriving at the limit
$\sigma\rightarrow 0$:

\beq
I_1=\frac{ig_{\mu\nu}}{(4\pi)^2}\left\{[m^2+q^2(x^2-x)]\ln\frac{m^2+q^2(x^2-x)}{\mu_3^2}+C_2\right\}
\label{5-7}
\eeq

with two arbitrary constant: $C_1=-\ln\mu_3^2$ and $C_2$. Combining
$I_1$ and $I_2$ together, we find:

\beq
\Pi_{\mu\nu}(q)=\frac{8ie^2}{(4\pi)^2}(q_{\mu}q_{\nu}-g_{\mu\nu}q^2)\int_0^1dx(x^2-x)
\ln\frac{m^2+q^2(x^2-x)}{\mu_3^2}-\frac{4ie^2}{(4\pi)^2}g_{\mu\nu}C_2
\label{5-8}
\eeq

The continuity equation of current induced in the vacuum
polarization \cite{21}

\beq
q^{\mu}\Pi_{\mu\nu}(q)=0
\label{5-9}
\eeq

is ensured by the factor $(q_{\mu}q_{\nu}-g_{\mu\nu}q^2)$. So we
set $C_2=0$. Consider the scattering between two electrons via the
exchange of a photon with momentum transfer $q\rightarrow0$.
Adding the contribution of $\Pi_{\mu\nu}(q)$ to the tree diagram
amounts to modify the charge square:

\beq e^2\longrightarrow e^2_R=Z_3e^2,\ \ \
Z_3=1+\frac{\alpha}{3\pi}(\ln\frac{m^2}{\mu_3^2}-\frac{q^2}{5m^2}+\cdots)
\label{5-10}
\eeq

As in Ref.\cite{8:b}, we will set $\mu_3=m$ so that at the Thomson
limit:$\lim_{q\rightarrow0}e_R^2=e^2$. However, for the purpose of
calculating Lamb shift (LS) below, the second term in the
parenthesis of $Z_3$ is important because for a bound state it
contributes a term of effective potential (adding to Coulomb
potential), called the Uehling potential (\cite{23},p.253):

\beq
-\frac{4\alpha^2}{15m^2}\delta(\mathbf r)
\label{5-11}
\eeq

\section{The Off-Mass-Shell Vertex Function}

Consider an electron (see Fig.2(d)) moving in a hydrogen atom,
its momentum changes from $p$ to $p'$ via the scattering by the
proton and an exchange of virtual photon with momentum $k$. The
FDI at one-loop level reads

\beq
\Lambda_{\mu}(p',p)=(-ie)^2\int\frac{d^4k}{(2\pi)^4}\frac{-i}{k^2}\gamma_{\nu}
\frac{i}{\not\!\!p'-\not\!\!k'-\mu}\gamma_{\mu}\frac{i}{\not\!\!p-\not\!\!k-\mu}\gamma^{\nu}
\label{6-1}
\eeq

However, different from \cite{8:b} and many other literatures, not
only the reduced mass $\mu$ (instead of $m$) of electron is used,
but also a new approach will be adopted. We assume that the
electron is moving off-mass-shell in the sense of (as in section
IV):

\beq
p^2=p'^2=\mu^2(1-\zeta)
\label{6-2}
\eeq

We still have

\beq
p'-p=q,\ \ \ p\cdot q=-\frac{1}{2}q^2
\label{6-3}
\eeq

Introducing Feynman parameters $u=x+y$ and $v=x-y$, we perform
a shift in the momentum integration:$k\rightarrow
K=k-(p+q/2)u-(q/2)v$, thus

\beq \Lambda_{\mu}=-ie^2[I_3\gamma_{\mu}+I_4]
\label{6-4}
\eeq

\bea
I_3&=&\int_0^1du\int_{-u}^udv\int\frac{d^4K}{(2\pi)^4}\frac{K^2}{(K^2-M^2)^3}\label{6-5}\\
M^2&=&[\mu^2(1-\zeta)-\frac{q^2}{4}]u^2+\frac{q^2}{4}v^2+\zeta\mu^2u\label{6-6}\\
I_4&=&\int_0^1du\int_{-u}^udv\int\frac{d^4K}{(2\pi)^4}\frac{A_{\mu}}{(K^2-M^2)^3}\label{6-7}\\
A_{\mu}\!\!\!&=&\!\!\!(4\!\!-\!\!4u\!\!-\!\!2u^2)\mu^2(1\!\!-\!\!\zeta)\gamma_{\mu}\!+\!2i(u^2\!\!-\!\!u)\mu
q^{\nu}\!\sigma_{\mu\nu}
\!-\!(2\!-\!2u\!+\!\frac{u^2}{2}\!-\!\frac{v^2}{2})q^2\gamma_{\mu}\!-\!(2\!+\!2u)v\mu
q_{\mu} \label{6-8} \ena

Set $K^2=K^2-M^2+M^2$, then $I_3=I'_3-\frac{i}{32\pi^2}$ and
$I'_3$ is only logarithmically divergent and so can be treated as
in previous sections, yielding:

\beq
I'_3=\frac{-i}{(4\pi)^2}\int_0^1du\int_{-u}^udv\ln\frac{M^2}{\mu_1^2}
\label{6-9}
\eeq

with $\mu_1$ an arbitrary constant.

However, unlike Ref.\cite{8:b} where the calculation was conducted on
the mass-shell, now the off-mass-shell integration in Eq.(\ref{6-9}) can be
performed in the approximation that $\frac{Q^2}{4\mu^2}\ll1$ and
$\zeta\ll1$ ($Q^2=-q^2$, $Q$ is the three-dimensional momentum
transfer) which will be enough to calculate the Lamb shift (LS).
Denoting

\beq
a=[\mu^2(1-\zeta)+\frac{Q^2}{4}]u^2+\zeta\mu^2u,\ \ \
b=\frac{Q^2}{4}
\label{6-10}
\eeq

we will perform the integration with respect to $v$ and $u$
rigorously:

\beq
\int_{-u}^udv\ln(a-bv^2)=2u[\ln\mu^2+\ln
u+\ln[(1-\zeta)u+\zeta]-4u+2\sqrt{\frac{4a}{Q^2}}\ln\frac{1+\sqrt{Q^2/4a}u}{1-\sqrt{Q^2/4a}u}
\label{6-11}
\eeq

Expanding the last term and keeping only up to the order of
$\zeta$ and $Q^2/4\mu^2$, we obtain

\beq
\int_0^1du\int_{-u}^udv\ln(a-bv^2)\simeq\ln\mu^2-1+\zeta+\frac{Q^2}{6\mu^2}(1-\zeta)
\label{6-12}
\eeq

To our great pleasure, throughout the evaluation of $I_4$, there
is no any infrared divergence which would appear in previous
literatures when integrating with respect to $u$ with lower limit
zero. To avoid the infrared divergence, \eg, in \cite{8:b}, a cutoff
was introduced at the lower limit. Now the infrared
divergence disappears due to the existence of off-mass-shell
parameter $\zeta$. For example, we encounter the following
integral, in which no cutoff is needed
($\lambda=(1-\zeta)+Q^2/4\mu^2\sim1$):

\beq
\int_0^1\frac{du}{u+\zeta/\lambda}=\frac{\zeta}{\lambda}-\ln\frac{\zeta}{\lambda}
\label{6-13}
\eeq

Hence, after elementary but tedious calculation, we find:

\beq\begin{array}{l}
\Lambda_{\mu}(p',p)=\frac{\alpha}{4\pi}[\frac{11}{2}-\ln\frac{\mu^2}{\mu_1^2}-3\zeta+4(1+\zeta)\ln\zeta]\gamma_{\mu}
+\frac{\alpha}{4\pi}\frac{Q^2}{\mu^2}\gamma_{\mu}(\frac{1}{6}+\frac{1}{2}\zeta+\frac{4}{3}\ln\zeta+2\zeta\ln\zeta)\\
+i\frac{\alpha}{4\pi}\frac{q^{\nu}}{\mu}\sigma_{\mu\nu}(1+3\zeta+2\zeta\ln\zeta)
\label{6-14}
\end{array}\eeq

\section{Calculation of Lamb Shift as an Off-Mass-Shell Effect at
One-Loop Level}

There are three parts in Eq.(\ref{6-14}). The first part in
combination with the vertex $\gamma_{\mu}$ at tree level provides
a renormalization factor as

\beq
Z_1^{-1}=1+\frac{\alpha}{4\pi}[\frac{11}{2}-\ln\frac{\mu^2}{\mu_1^2}-3\zeta+4(1+\zeta)\ln\zeta]
\label{7-1}
\eeq

Further combination with $Z_2$ in Eq.(\ref{4-10}) and $Z_3$ in
Eq.(\ref{5-10}) leads to a renormalized charge (at one-loop level,
see Fig.2):

\beq
e_R=\frac{Z_2}{Z_1}Z_3^{1/2}e
\label{7-2}
\eeq

However the Ward identity implies that \cite{21,22,23} \

\beq
Z_1=Z_2
\label{7-3}
\eeq

Therefore

\beq
\alpha_R=\frac{e_R^2}{4\pi}=Z_3\alpha
\label{7-4}
\eeq

Note that Ward identity holds not only for an electron on the
mass-shell, but also for off-mass-shell case. Hence for every
bound state in hydrogenlike atom with a definite value of $\zeta$
($Z_1$ and $Z_2$ are functions of $\zeta$), the arbitrary constant
$\mu_1$ in Eq.(\ref{7-1}) plays a flexible role to guarantee the
validity of Eq.(\ref{7-3}) (other two constants $\mu_2$ and $\mu_3$
had been fixed in Eq.(\ref{4-13}) and (\ref{5-10}) respectively). For further discussion, see section IX.

The second part of Eq.(\ref{6-14}) contains $Q^2\gamma_{\mu}$. Just
like the Uehling potential in Eq.(\ref{5-10}) (with $q^2=-Q^2$), it
contributes an effective potential of $\delta$ function type as

\beq
\frac{\alpha^2}{\mu^2}[-\frac{1}{6}-\frac{1}{2}\zeta-\frac{4}{3}\ln\zeta-2\zeta\ln\zeta]\delta(\mathbf
r)
\label{7-5}
\eeq

Finally, the third part of Eq.(\ref{6-14}) amounts to a modification
of electron magnetic moment in the atom, the gyromagnetic ratio of
electron reads:

\beq
g=2[1+\frac{\alpha}{2\pi}(1+3\zeta+2\zeta\ln\zeta)]
\label{7-6}
\eeq

We will call the anomalous part of magnetic
moment $a=\frac{\tilde{\alpha}}{2\pi}$, $\tilde{\alpha}=\alpha(1+3\zeta+2\zeta\ln\zeta)$. The radiative correction
on the magnetic moment of an electron has two consequences. One is
a modification to the L-S coupling in a hydrogenlike atom (with
charge number $Z$) \cite{21,22}:

\beq
H_{LS}^{rad}=2(\frac{\tilde{\alpha}}{2\pi})\frac{\alpha
Z}{4\mu^2r^3}{\mathbf \sigma\cdot L}
\label{7-7}
\eeq

Here the electron mass
has been modified from $m$ (see, \eg, \cite{15}) to $\mu$ which
can be derived from the reduced Dirac equation.

Another consequence of anomalous magnetic moment of electron
exhibits itself as an additional potential of $\delta$ function
type like Eq.(\ref{5-11})\cite{21,22}

\beq
\frac{Z\alpha\tilde{\alpha}}{2\mu^2}\delta(\mathbf r)
\label{7-8}
\eeq

Note that Eqs.(\ref{7-7}) and (\ref{7-8}) are only effective to states
with $L\not=0$ and $S$ state with $L=0$ respectively.

Adding the results of Eqs.(\ref{7-7}), (\ref{7-8}) and the sum of Eqs.(\ref{5-11}) and (\ref{7-5})
multiplied by $Z$ together to get all radiative corrections (at one-loop level) on
electron in the hydrogenlike atom, then we get the effective potential as

\beq
\begin{array}{ll}
V_{eff}^{rad}&=\frac{Z\alpha^2}{\mu^2}[-\frac{4}{3}\ln\zeta-\frac{1}{2}\zeta-2\zeta\ln\zeta
-\frac{1}{6}-\frac{4}{15}\frac{\mu^2}{m^2}+\frac{1}{2}(1+3\zeta+2\zeta\ln\zeta)
]\delta(\mathbf r)\\
&+\frac{Z\alpha^2}{4\pi\mu^2r^3}(1+3\zeta+2\zeta\ln\zeta)
{\mathbf \sigma\cdot L}\\
&\simeq\frac{Z\alpha^2}{\mu^2}[-\frac{4}{3}\ln\zeta+\frac{1}{15}+\zeta-\zeta\ln\zeta]\delta(\mathbf
r)+\frac{Z\alpha^2}{4\pi\mu^2r^3}(1+3\zeta+2\zeta\ln\zeta){\mathbf \sigma\cdot L}
\end{array}
\label{7-9}
\eeq

where we take $\mu^2/m^2\approx1$ in the Uehling potential
to make the formula simpler for a semi-quantitative calculation.
Eq.(\ref{7-9}) leads to the energy modification of a bound state
(with quantum numbers $n,l,j$) in a hydrogenlike atom:

\beq
\delta({\mathbf r})\longrightarrow
|\psi_{ns}(0)|^2=\frac{Z^3\alpha^3}{\pi n^3}\mu^3,\ \ \ (l=0)
\label{7-10}
\eeq

\beq
\Delta E^{rad}=\Delta E^{rad}(ns)+\Delta E^{rad}_{LS}
\label{7-11}
\eeq

\beq \Delta E^{rad}(ns)=\frac{Z^4\alpha^3}{\pi
n^3}R_y[\frac{8}{3}\ln\frac{1}{\zeta}+\frac{2}{15}+2\zeta(1-\ln\zeta)]\delta_{l0}
\label{7-12}
\eeq

\beq
\Delta E^{rad}_{LS}=\frac{Z^4\alpha^3}{\pi
n^3}R_y\frac{1+\zeta(3+2\ln\zeta)}{l(2l+1)(l+1)}\left\{
\begin{array}{ll}
&l,\ \ \ \ \ \ (j=l+1/2)\\
-&(l+1),\ \ \ (j=l-1/2)
\end{array}
\right.
\label{7-13}
\eeq

where

\beq
R_y=\frac{1}{2}\alpha^2\mu=\frac{\mu}{m}R_\infty
\label{7-14}
\eeq

\section{Energy-Level Difference in Hydrogenlike Atom: Theory vs.
Experiment}

We will study some energy-level differences near the ground state
of hydrogenlike atoms, where precise experimental data are
available. Theoretically, the energy level is fixed primarily by
the formula derived from the reduced Dirac equation (RDE), \ie,
Eq.(\ref{1-1}) with $m_e$ substituted by $\mu_A$ where the subscript $A$
refers to atom $H$, $D$ or $He^+$, \etal.:

\bea
E_A^{RDE}&=&\mu_A[f(n,j)-1]=\frac{m_e}{1+b_A}[f(n,j)-1]\nonumber\\
&=&\frac{1}{1+b_A}(1.2355897\times10^{20})[-\frac{(Z\alpha)^2}{2n^2}-\frac{(Z\alpha)^4}{3n^3}(\frac{1}{j+1/2}-
\frac{3}{4n})-\cdots]\ \ Hz
\label{8-1}
\ena

Further recoil corrections Eq.(\ref{1-6}) derived by previous
authors will be divided into two terms:

\bea
\Delta E_A^{recoil-1}(n,j)&=&-\frac{\mu^2_A}{2M_A}[f(n,j)-1]^2=-\frac{m_eb_A}{2(1+b_A)^3}[f(n,j)-1]^2
\label{8-2}\\
\Delta E_A^{recoil-2}(n,j,l)&=&\frac{(Z\alpha)^4\mu^3_A}{2n^3m_N^{(A)^2}}
(\frac{1}{j+\frac{1}{2}}-\frac{1}{l+\frac{1}{2}})(1-\delta_{l0})
\label{8-3}
\ena

Next comes the radiative correction calculated by QED at one-loop
level, Eq.(\ref{7-11}):

\beq \Delta
E_A^{rad}(n,j,l)=\frac{1}{1+b_A}\frac{Z^4}{n^3}(\frac{\alpha^3}{\pi}R_{\infty})
[(-\frac{8}{3}\ln\zeta+\frac{2}{15}+2\zeta(1-\ln\zeta))\delta_{l0}
+\frac{1+\zeta(3+2\ln\zeta)}{2l+1}C_{jl}(1-\delta_{l0})]
\label{8-4}
\eeq

where

\beq
C_{jl}=\left\{
\begin{array}{ll}
&\frac{1}{l+1},\ \ \ j=l+\frac{1}{2}\\
&-\frac{1}{l},\ \ \  j=l-\frac{1}{2}
\end{array}
\right.
\label{8-5}
\eeq

Finally, the finite nucleus size (NS) with radius $r_N^{(A)}$
brings a correction \cite{10}:

\beq
\begin{array}{ll}
\Delta E_A^{NS}(n,j)&=\frac{4}{3}(\frac{\mu_A}{m_e})^3\frac{Z^4}{n^3}(\frac{r_N^{(A)}}{a_{\infty}})^2R_{\infty}\delta_{l0}\\
&=(\frac{1}{1+b_A})^3\frac{Z^4}{n^3}(4.386454987\times10^7)[\frac{r_N^{(A)}(fm)}{5.2917725}]^2\delta_{l0}\
\ Hz
\end{array}
\label{8-6}
\eeq

As explained in Eq.(\ref{4-16}) with Eq.(\ref{4-17}), the value of
off-mass-shell parameter $\zeta$ in Eq.(\ref{8-4}) can be calculated
from the electron self-energy at one-loop level:

\beq
\frac{Z^2\alpha}{n^2}=\frac{1}{2\pi}\frac{(\zeta^{<S>}-2\zeta^{<S>}\ln\zeta^{<S>})}{1+\alpha/3\pi}
\label{8-7}
\eeq

where the superscript $<S>$ refers to "self-energy". However, we
may derive the value of $\zeta$ in an alternative way. Divide the
square average of four-dimensional momentum $p$ into two parts:

\beq
<p^2>=<E^2>-<{\mathbf p}^2>
\label{8-8}
\eeq

where

\beq
<E^2>=E^2=(\mu-B)^2\simeq\mu^2-2\mu B,
\label{8-9}
\eeq

since the binding energy

\beq
B=\frac{Z^2\alpha^2}{2n^2}\mu\ll\mu
\label{8-10}
\eeq

The square average of three-dimensional momentum $\mathbf p$,
$<{\mathbf p}^2>$, can be evaluated by the Virial theorem (\eg,
\cite{15}). In a Coulomb field, an electron has potential energy
$V=-\frac{Ze^2}{4\pi r}$ and kinetic energy
$T=\frac{1}{2\mu}{\mathbf p}^2$. Then

\bea
<{\mathbf p}^2>&=&2\mu<T>=2\mu[-B-<V>]=2\mu B\nonumber\\
<p^2>&=&\mu^2-4\mu B=\mu^2(1-\frac{4B}{\mu})
\label{8-11}
\ena

Comparing Eq.(\ref{8-11}) with $<p^2>=\mu^2(1-\zeta^{<V>})$, we find

\beq
\zeta^{<V>}=\frac{4B}{\mu}=\frac{2Z^2\alpha^2}{n^2}
\label{8-12}
\eeq

where the superscript $<V>$ refers to "Virial theorem". Table 1
gives the values of $\zeta^{<S>}$ and $\zeta^{<V>}$ with their
logarithm values as well as two kinds of "average", $\zeta^{<S+V>}=\frac{1}{2}(\zeta^{<S>}+\zeta^{<V>})$
and $\zeta^{<SV>}=\sqrt{\zeta^{<S>}\zeta^{<V>}}$, to be used in
Eq.(\ref{8-4}).

\vskip 0.5cm
\small
\renewcommand\arraystretch{1.3}
\begin{small}\hspace*{-16mm}\begin{tabular}{|c|c|c|c|c|c|c|c|c|}
\multicolumn{8}{c}{Table 1. Off-mass-shell parameter $\zeta$ and $\ln\zeta$}\\[5pt]
  \hline
  $\frac{Z^2}{n^2}$ & $\zeta^{<S>}\times 10^4$ &-$\ln\zeta^{<S>}$ & $\zeta^{<V>}\times 10^6$ & -$\ln\zeta^{<V>}$ &
  $\zeta^{<S+V>}\times 10^5$ & -$\ln\zeta^{<S+V>}$ & $\zeta^{<SV>}\times 10^5$ & $-\ln\zeta^{<SV>}$ \\
  \hline
  $\frac{1}{16}$ & $1.546093458$ & $8.77461$ & $\frac{\alpha^2}{8}=6.6564192$
   &11.91992886 & $8.0632$ & 9.425609 & 3.2080284 & 10.34727 \\
   \hline
  $\frac{1}{4}$ & $7.446539697$ & 7.20259 & $\frac{\alpha^2}{2}=26.6256771$
  & 10.5336345 & $38.5639$ & 7.860609 & 14.0808 & 8.86816225 \\
   \hline
 1 & $37.73719345$ & 5.57969 & $2\alpha^2=106.502$ &9.147340142 &
 $194.011$ & 6.2450103 & 63.39626 & 7.36351521 \\
  \hline
\end{tabular}\end{small}
\vspace{0.5cm}
\normalsize

Now we are in a position to discuss a number of cases:

(a) The so-called classic Lamb shift of hydrogen atom was measured
experimentally as \cite{10}:

\beq
L_H^{exp}(2S-2P)\equiv E_H(2S_{1/2})-E_H(2P_{1/2})=1057.845\
\ MHz
\label{8-13}
\eeq

Theoretically, in this case ($b_H=5.446170255\times10^{-4},\ r_N^H=r_p=0.862 fm$),
Eqs.(\ref{8-1}) and (\ref{8-2}) make no contributions while
Eqs.(\ref{8-3}) and (\ref{8-6}) only contribute

\beq
\Delta E_H^{recoil-2}(2S_{1/2}-2P_{1/2})=-E_H^{recoil-2}(2,1/2,1)=-2.16156\ \ kHz
\label{8-14}
\eeq

and

\beq
\Delta E_H^{NS}(2S-2P)=0.14525347\ \ MHz
\label{8-15}
\eeq

respectively. The dominant contribution comes from Eq.(\ref{8-4}).
If using $\zeta^{<S>}$, we obtain

\beq\begin{array}{l}
\Delta
E_H^{Rad<S>}(2S-2P)=\frac{1}{1+b_H}\frac{1}{8}(4.06931316\times10^8)
[-\frac{8}{3}\ln\zeta^{<S>}+\frac{7}{15}+3\zeta^{<S>}-\frac{4}{3}\zeta^{<S>}\ln\zeta^{<S>}]\\
=1000.6567\
MHz \label{8-16} \end{array}\eeq

If we use another three values of $\ln\zeta$ in Table 1, we get

\bea \Delta E_H^{Rad<V>}(2S-2P)&=&1451.7912\ \ MHz \label{8-17}\\
\Delta E_H^{Rad<S+V>}(2S-2P)&=&1089.6513\ \ MHz \label{8-18}\\
\Delta E_H^{Rad<SV>}(2S-2P)&=&1226.0871\ \ MHz
 \label{8-19}
\ena

It seems that Eq.(\ref{8-16}) is smaller whereas Eq.(\ref{8-17}) too
large. So as an empirical rule in our semiquantitative
calculation, we may use Eq.(\ref{8-18}) to get

\beq L_H^{theor.}(2S-2P)= 1089.651+0.145-0.002=1089.794\ \ MHz
\label{8-20} \eeq

which is larger than Eq.(\ref{8-13}) by $3\%$.

(b) The Lamb shift of $He^+$ atom has been measured as (quoted
from \cite{27}):

\beq L_{He^+}^{exp}(2S-2P)= 14041.13(17)\ \ MHz \label{8-21} \eeq

Similar to the case of hydrogen atom but with $Z=2$ and
$b_{He^+}=\frac{m_e}{m_{\alpha}}=0.0001371$, we find

\bea \Delta E_{He^+}^{Rad<S>}(2S-2P)&=&1.252680693\times10^{10}\ \ Hz\nonumber\\
\Delta E_{He^+}^{Rad<V>}(2S-2P)&=&2.023083608\times10^{10}\ \ Hz\nonumber\\
\Delta E_{He^+}^{Rad<S+V>}(2S-2P)&=&1.369980830\times10^{10}\ \ Hz\nonumber\\
\Delta E_{He^+}^{Rad<SV>}(2S-2P)&=&1.636521214\times10^{10}\ \ Hz
 \label{8-22}
\ena

As in the case of $H$ atom, we take the $<S+V>$ scheme and add

\bea \Delta E_{He^+}^{recoil-2}(2S-2P)&=&-2.165\ \ kHz \label{8-23}\\
\Delta E_{He^+}^{NS}(2S-2P)&=&4.514\ \ MHz \label{8-24} \ena

($r_{\alpha}\simeq1.2fm$), to find the theoretical value:

\beq L_{He^+}^{theor.}(2S-2P)= 13704.220\ \ MHz \label{8-25} \eeq

which is smaller than Eq.(\ref{8-21}) by $2.41\%$.

(c) The following energy-level difference is related to the "hyper
Lamb shift (HLS)" \cite{10}:

\beq \Delta_H^{exp}\equiv
E_H(4S)-E_H(2S)-\frac{1}{4}[E_H(2S)-E_H(1S)]=4797.338(10)\ \ MHz
\label{8-26} \eeq

Theoretically, now Eq.(\ref{8-1}) makes the main contribution:

\beq \Delta
E_H^{RDE}[(4S)-\frac{5}{4}(2S)+\frac{1}{4}(1S)]=3923.95\ \ MHz
\label{8-27} \eeq

(The notation in parenthesis is self-evident). Eq.(\ref{8-2}) and Eq.(\ref{8-4})
contribute

\beq \Delta E_H^{recoil-1}[(4S)-\frac{5}{4}(2S)+\frac{1}{4}(1S)]=
-4.186\ \ MHz \label{8-28} \eeq

and

\bea \Delta E_H^{Rad<S>}&=&451.229097\ \ MHz\nonumber\\
\Delta E_H^{Rad<S+V>}&=&529.288296\ \ MHz\nonumber\\
\Delta E_H^{Rad<SV>}&=&675.907131\ \ MHz\nonumber\\
\Delta E_H^{Rad<V>}&=&903.266275\ \ MHz
 \label{8-29}
\ena

respectively. Adding a small contribution from Eq.(\ref{8-6})

\beq \Delta E_H^{NS}[(4S)-\frac{5}{4}(2S)+\frac{1}{4}(1S)]=
0.1270967854\ \ MHz \label{8-30} \eeq

we get
\bea
\Delta_H^{Theor.<S>}&=&4371.120197\ \ MHz\nonumber\\
\Delta_H^{Theor.<S+V>}&=&4449.179396\ \ MHz\nonumber\\
\Delta_H^{Theor.<SV>}&=&4595.798231\ \ MHz\nonumber\\
\Delta_H^{Theore.<V>}&=&3923.95-4.186+903.266275+0.1271=4823.1574\ \ MHz \label{8-31}
\ena

The $<V>$ scheme is only larger than Eq.(\ref{8-26}) by $0.54\%$.
All other schemes would be too small. So we guess that
for $S$ states $<V>$ scheme is better than $<S>$ scheme.

(d) The following energy-level difference was also measured as
\cite{10}:

\beq {\Delta'}_H^{exp}\equiv
E_H(4D_{5/2})-E_H(2S)-\frac{1}{4}[E_H(2S)-E_H(1S)]=6490.144(24)\ \
MHz \label{8-32} \eeq

Theoretically, Eq.(\ref{8-1}) also makes the main contribution:

\beq \Delta
E_H^{RDE}[(4D_{5/2})-\frac{5}{4}(2S)+\frac{1}{4}(1S)]=5747.92\ \
MHz \label{8-33} \eeq

while

\beq \Delta
E_H^{recoil-1}[(4D_{5/2})-\frac{5}{4}(2S)+\frac{1}{4}(1S)]=-4.18611\
\ MHz \label{8-34} \eeq

\beq \Delta
E_H^{recoil-2}[(4D_{5/2})]=\alpha^4m_e(5.446170255\times10^{-4})^2(\frac{1}{3}-\frac{2}{5})=-6.9283\
\ kHz \label{8-35} \eeq

are all small, we will have

\bea {\Delta'}
E_H^{rad<S>}[(4D_{5/2})-\frac{5}{4}(2S)+\frac{1}{4}(1S)]&=&
302.088631\ \ MHz\nonumber\\
{\Delta'}E_H^{rad<V>}[(4D_{5/2})-\frac{5}{4}(2S)+\frac{1}{4}(1S)]&=&
700.843464\ \ MHz\nonumber\\
{\Delta'}E_H^{rad<S+V>}[(4D_{5/2})-\frac{5}{4}(2S)+\frac{1}{4}(1S)]&=&
369.124660\ \ MHz\nonumber\\
{\Delta'}E_H^{rad<SV>}[(4D_{5/2})-\frac{5}{4}(2S)+\frac{1}{4}(1S)]&=&
500.131264\ \ MHz \label{8-36} \ena

Finally, the nucleus size effect gives

\beq
{\Delta'}E_H^{NS}[(4D_{5/2})-\frac{5}{4}(2S)+\frac{1}{4}(1S)]=
11.62027752\times10^5(\frac{1}{4}-\frac{5}{4}\times\frac{1}{8})=0.10894\
\ MHz \label{8-37} \eeq

In sum, we have

\bea
{\Delta'}_H^{<S>}&=&{\Delta'}E_H^{RDE}+{\Delta'}E_H^{recoil-1}+{\Delta'}E_H^{recoil-2}+{\Delta'}
E_H^{rad<S>}+{\Delta'} E_H^{rad<NS>}=6045.925\ \ MHz\nonumber\\
{\Delta'}_H^{<V>}&=&6444.679\ \ MHz\nonumber\\
{\Delta'}_H^{<S+V>}&=&6112.961\ \ MHz\nonumber\\
{\Delta'}_H^{<SV>}&=&6243.967\ \ MHz \label{8-38}
\ena

which are smaller than the experimental value (\ref{8-32}) by
$6.8\%,\  0.7\%,\  5.8\%$ and $3.8\%$ respectively.
.

(e) Experimentally, the combination of Eq.(\ref{8-26}) with
Eq.(\ref{8-32}) yields:

\beq {\Delta''}_H^{exp}\equiv E(4D_{5/2})-E(4S_{1/2})=1692.806\ \
MHz \label{8-39} \eeq

Then, theoretically, we have

\bea
{\Delta''}_H^{RDE}(4D_{5/2}-4S)&=&1.823886903\times10^9\ \ Hz \label{8-40}\\
{\Delta''}_H^{recoil-1}(4D_{5/2}-4S)&=&1.1008\ \ Hz \label{8-41}\\
{\Delta''}_H^{recoil-2}(4D_{5/2}-4S)&=&-6.9283\ \ kHz \label{8-42}\\
{\Delta''}_H^{NS}(4D_{5/2}-4S)&=&-0.0181605862\ \ MHz \label{8-43}
\ena

and

\bea
{\Delta''}_H^{rad<S>}(4D_{5/2}-4S)&=&-149.1404661\ \ MHz \label{8-44}\\
{\Delta''}_H^{rad<V>}(4D_{5/2}-4S)&=&-202.4228107\ \ MHz \label{8-45}\\
{\Delta''}_H^{rad<S+V>}(4D_{5/2}-4S)&=&-160.1636366\ \ MHz \label{8-46}\\
{\Delta''}_H^{rad<SV>}(4D_{5/2}-4S)&=&-175.7758676\ \ MHz \label{8-47}
\ena

Altogether, we have

\bea
{\Delta''}_H^{theore.<S>}(4D_{5/2}-4S)&=&1674.721349\ \ MHz\nonumber\\
{\Delta''}_H^{theore.<S+V>}(4D_{5/2}-4S)&=&1663.716339\ \ MHz\nonumber\\
{\Delta''}_H^{theore.<SV>}(4D_{5/2}-4S)&=&1648.104108\ \ MHz\nonumber\\
{\Delta''}_H^{theore.<V>}(4D_{5/2}-4S)&=&1621.439105\ \ MHz
\label{8-48}
\ena

which are smaller than Eq.(\ref{8-39}) by $1.1\%,\ 1.7\%,\ 2.6\%$ and $4.2\%$ respectively.

(f) It's time to go back to the precision data of $2S-1S$
transition in hydrogen atom as discussed in section II. Rewrite
Eq.(\ref{2-1}) as (see also \cite{43}):

\beq \Delta E_H^{exp}(2S-1S)=2.46606141318734\times10^{15}\ \ Hz
\label{8-49} \eeq

Theoretically, we have [see Eq.(\ref{2-4})]:

\beq \Delta E_H^{RDE}(2S-1S)=2.466067984\times10^{15}\ \ Hz
\label{8-50} \eeq

\beq \Delta E_H^{recoil-1}(2S-1S)=22.32598676\ \ MHz
\label{8-51}
\eeq

\bea \Delta E_H^{rad<S>}(2S-1S)&=&-5142.081146\ \ MHz\nonumber\\
\Delta E_H^{rad<S+V>}(2S-1S)&=&-5765.958928\ \ MHz\nonumber\\
\Delta E_H^{rad<SV>}(2S-1S)&=&-6835.535314\ \ MHz\nonumber\\
\Delta E_H^{rad<V>}(2S-1S)&=&-8541.095068\ \ MHz
 \label{8-52}
\ena

\beq
\Delta E_H^{NS}(2S-1S)=11.62027752\times10^5(\frac{1}{8}-1)=-1.016774283\
\ MHz
 \label{8-53}
\eeq

If taking the value of $\Delta E_H^{rad}(2S-1S)$, we get

\bea
\Delta E_H^{theore.<S>}(2S-1S)&=& 2.466062836\times10^{15} \ \ Hz\nonumber\\
\Delta E_H^{theore.<S+V>}(2S-1S)&=& 2.466062239\times10^{15} \ \ Hz\nonumber\\
\Delta E_H^{theore.<SV>}(2S-1S)&=& 2.466061169\times10^{15} \ \ Hz\nonumber\\
\Delta E_H^{theore.<V>}(2S-1S)&=& 2.466059464\times10^{15}\ \ Hz \label{8-54}
\ena

They are larger than Eq.(\ref{8-49}) by $1450\ MHz,\ 826\ MHz$ and smaller than Eq.(\ref{8-49})
by $244\ MHz,\ 1949\ MHz$ respectively. Or, their discrepancies are
$+5.9\times10^{-7},\ +3.3\times10^{-7},\ -1.0\times10^{-7},\ -7.9\times10^{-7}$, respectively.
This discrepancy is basically stemming from
the uncertainty in the calculation of $\Delta E_H^{rad}(2S-1S)$.

(g) Let us turn to the isotope-shift of $2S-1S$ transition.
Rewrite Eq.(\ref{2-2}) as

\beq \Delta E_{D-H}^{exp}(2S-1S)=6.70994337\times10^{11}\ \ Hz
\label{8-55} \eeq

Theoretically, rewrite Eqs.(\ref{2-6}) and (\ref{2-7}) as

\beq \Delta E_{D-H}^{RDE}(2S-1S)=6.7101527879\times10^{11}\ \ Hz
\label{8-56} \eeq

and

\beq \Delta E_{D-H}^{recoil-1}(2S-1S)=-11.176\ \ MHz \label{8-57}
\eeq

\bea \Delta E_{D-H}^{rad<S>}(2S-1S)&=&-1.399158\ \ MHz\nonumber\\
\Delta E_{D-H}^{rad<V>}(2S-1S)&=&-2.324028\ \ MHz\nonumber\\
\Delta E_{D-H}^{rad<S+V>}(2S-1S)&=&-1.568915\ \ MHz\nonumber\\
\Delta E_{D-H}^{rad<SV>}(2S-1S)&=&-1.859945\ \ MHz
 \label{8-58}
\ena

\beq \Delta E_{D-H}^{NS}(2S-1S)=-5.11384949\ \ MHz \label{8-59}
\eeq

Altogether, we find [using $<V>$ scheme in Eq.(\ref{8-58})]:

\beq \Delta
E_{D-H}^{theore.<V>}(2S-1S)=6.709966701\times10^{11}\ \ Hz
\label{8-60} \eeq

which is larger than Eq.(\ref{8-55}) by $2.333\ MHz$ or only
$3.5\times10^{-6}$. Evidently, even Eq.(\ref{8-56}) solely
deviates from Eq.(\ref{8-55}) by $3\times10^{-5}$ only. And as
expected, the different schemes for $\Delta E_{D-H}^{rad}(2S-1S)$
have little influence on the theoretical value, because any one of
Eq.(\ref{8-58}) is much smaller than the nucleus size effect
Eq.(\ref{8-59}) ($r_N^D=r_d=2.115 fm$).

(h) Finally, the so-called absolute Lamb-shift of $1S$ state in
hydrogen atom was determined by Weitz \etal \cite{10}
from the measured value Eq.(\ref{8-26}) or (\ref{8-32}). In our
notation, using Eq.(\ref{8-32}), we will write it as follows:

\beq
\begin{array}{ll}
L_H(1S)=&4\{{\Delta'}_H^{exp}-\Delta
E_H^{RDE}[(4D_{5/2})-\frac{5}{4}(2S)+\frac{1}{4}(1S)]-\Delta
E_H^{recoil-1}[(4D_{5/2})-\frac{5}{4}(2S)+\frac{1}{4}(1S)]\\
&-\Delta
E_H^{recoil-2}(4D_{5/2})+\frac{5}{4}L_H(2S)-L_H(4D_{5/2})\}
\end{array}
\label{8-61} \eeq

Here the Lamb shift of $2S$ state $L_H(2S)$ can be determined from
the experimental value of Eq.(\ref{8-13}) with $L_H(2P_{1/2})$
being calculated from Eq.(\ref{8-4}):

\beq L_H(2S)=L_H^{exp}(2S-2P)-\Delta
E_H^{recoil-2}(2S-2P_{1/2})+L_H(2P_{1/2})=1040.901\ \ MHz
\label{8-62} \eeq

And $L_H(4D_{5/2})$ can also be calculated from Eq.(\ref{8-4}), so

\beq L_H(1S)=8188.478\ \ MHz \label{8-63} \eeq

which is in agreement with $8172.874(60)\ MHz$ given by \cite{10}
within an accuracy $\lesssim 0.2\%$. If we use Eq.(\ref{8-26}) to
derive $L_H(1S)$, we would have to calculate $L_H(4S)$ which is
much larger than $L_H(4D_{5/2})$ and its derivation from
Eq.(\ref{8-4}) seems not reliable. Similarly, the theoretical
value of $L_H(1S)$ turns out to be

\beq L_H^{theore.}(1S)=\Delta E_H^{rad}(1S)+\Delta E_H^{NS}(1S)
\label{8-64} \eeq

with $\Delta E_H^{NS}(1S)=0.14525347\ MHz$. However, the value of $\Delta
E_H^{rad}(1S)$ strongly depends on the scheme we used in
Eq.(\ref{8-4}), which must
be narrowed in a high-loop calculation.
The theoretical prediction was given in \cite{27} as:
\begin{equation}\label{8-65}
    L_H^{theor.}(1S)=8172754(14)(32)\ kHz
\end{equation}
Further discussions can be found in Refs. \cite{5,44,45}.

\section{Summary and Discussion}

The remarkable progress of the experimental research on
energy-level differences in hydrogenlike atoms has been making
this field an ideal theoretical laboratory for physics:

(a) The inevitable and successful use of reduced Dirac equation
(RDE) to hydrogenlike atoms, especially to the isotope-shift of
$2S-1S$ transition as reflected by Eqs.(\ref{8-55}) through
(\ref{8-60}), is by no means an accidental fortune. It implies
that the argument in section III for introducing RDE, Eq.(\ref{3-21}),
is correct to a high accuracy. In particular, the basic principle of
invariance under space-time inversion Eq.(\ref{3-22}) (with original mass $m$) could remain
valid even for a noninertial frame. This implication has a far-reaching consequence that a
generalization at the above symmetry to a localized curved
space-time may be served as a possible road to quantize the
general theory of relativity \cite{16}.

However, there are two realizations of potential $V$ under the
space-time inversion, Eq.(\ref{3-15})("scalar" type) and
Eq.(\ref{3-22})("vector" type). While Eq.(\ref{3-22}) does
dominant in an atom like $H$ with $m_p\gg m_e$, the remaining
discrepancy of $2.333\ MHz$ between theory and experiment
[Eq.(\ref{8-60}) versus Eq.(\ref{8-55})] strongly hints that an
important and subtle effect had been ignored. (To consider the contribution of
the deuteron polarizability merely accounts for about $20\ kHz$
\cite{6}). We think what neglected must be a tiny excitation of antiparticle
field in the nucleus due to its interaction with electron in the CMCS. So when we reduce the degrees of freedom
of two-body system from two to
one, the RDE should be modified
to take account of the tiny mixture of "scalar" potentials (see the page note after Eq.(\ref{3-22})).
We don't know how to improve $RDE$ yet.
However, an experimental evidence for the above conjecture could be the
following prediction: The discrepancy between present theory (with RDE) and
experiment must be smaller for the isotope shift in $2S-1S$
transition of atoms $^4He$ and $^3He$ than that of atoms $H$ and $D$.

Recently, by using Dirac's method, Marsch rigorously solved the
hydrogen atom as a two-Dirac particle system bound by Coulomb
force \cite{34}. His solutions are composed of positive and
negative pairs, corresponding respectively to hydrogen and
anti-hydrogen as expected. However, surprisingly, in the hydrogen
spectrum, besides the normal type-1 solution with reduced mass $\mu$, there is another
anomalous type$-2$ solution with energy levels:  ${E'}_n=
Mc^2-2\mu c^2+\frac{1}{2}\mu c^2(\frac{\alpha}{n})^2+\cdots\
(n=1,2,\ldots)$ and "strange enough, the type$-2$ ground state
$(n=1)$ does not have lowest energy but the continuum
$(n=\infty)$". In our opinion, based on what we learnt from the
Dirac equation and RDE, these anomalous solutions imply a positron
moving in the field of proton. So all discrete states with energy
${E'}_n$ are actually unbound, they should be and can be ruled out
in physics either by the "square integrable condition" or the
"orthogonality criterion" acting on their rigorous wave functions
(for one-body Dirac equation, see \cite{35}, also p.$28-31$, $50$ of \cite{36}). On the other
hand, all continuum states ($n=\infty$) with energies lower than
$Mc^2-2\mu c^2$ correspond to scattering wave functions with
negative phase shifts , showing the repulsive force between
positron and proton. (see \cite{37}, section 1.5 in \cite{36} or
section 9.5 of \cite{15}). Marsch's discovery precisely reflects
two things: (a) the negative energy state of a particle just
describes its antiparticle state. (b) The Coulomb potential allows
a complete set of solutions comprising of two symmetric
sectors,hydrogen and antihydrogen.In the hydrogen sector, the
proton remains unchanged regardless of the changing process of
electron into positron under the Coulomb interaction.

The above particle-antiparticle symmetry (including Eq.(\ref{3-22}) showing the unequal treatment between electron and
nucleus), together with the parity symmetry, is hidden in the Dirac's
four-component theory in covariant form so they were overlooked to some extent in the past. The advantage
or flexibility of two-component noncovariant form of Dirac equation or RDE (as discussed in this
paper) lies in the fact that the above two symmetries become accurate and so easily to be extended
(or violated) in an explicit manner. For completeness, let us stress again that for antiparticle, one should
use the momentum and energy operators being ${\mathbf p}_c=i\nabla$ and $E_c= -i\frac{\partial}{\partial t}$
versus ${\mathbf p}=-i\nabla$ and $E=i\frac{\partial}{\partial t}$ for particle as required by the
space-time inversion symmetry. The historical mission of the conception to imagine the positron as a
"hole" in the sea of negative energy electrons is already over. Since the CPT invariance had been further verified \cite{39},
the relation between a particle $|a\rangle$ and its antiparticle $|\bar{a}\rangle$ is well-established as: \footnotemark[4]
\footnotetext[4]{To our knowledge, the correct definition, Eq.(\ref{167}), was first given by T. D. Lee and C. S. Wu at
Ann. Rev. Nucl. Sci. {\bf 15}, 381(1965). See also G. J. Ni at J. Fudan Univ. (Natural Science) No.3-4, 125(1974).}

\begin{equation}\label{167}
|\bar{a}\rangle=CPT|a\rangle
\end{equation}
with their wave-functions (in free motion) being respectively:
\begin{equation}\label{awave}
\langle{\bf x},t|a\rangle\sim \exp[\frac{i}{\hbar}({\bf p}\cdot{\bf x}-Et)]
\end{equation}
\begin{equation}\label{abar}
\langle{\bf x},t|\bar{a}\rangle\sim \exp[-\frac{i}{\hbar}({\bf p}\cdot{\bf x}-Et)]
\end{equation}
Note that in Eqs.(\ref{awave}) and (\ref{abar}), they have the same momentum $\bf p$ and positive energy $E$.
Either a newly defined space-time inversion (${\bf x}\to -{\bf x},\,t\to -t$)
or a simple change of $i\to -i$ will transform Eq.(\ref{awave}) into Eq.(\ref{abar}) (or vice versa).

(b) Throughout this paper, the electron bound in an atom
is just treated like a stationary "ball" with nucleus at its center and
having a (Bohr) radius ($\sim 1/\alpha m_e$). However, it is in an
off-mass-shell state (In some sense, our atom model is just the opposite to J. J. Thomson's atom model
100 years ago). In fact, the electron's mass is reduced
suddenly from $m_e$ to $\mu$ in the RMCS when it is captured by a
nucleus at the far remote orbit with quantum number
$n\longrightarrow\infty$ and further reduced to
$\mu+\delta\mu\simeq\mu-\frac{Z^2\alpha^2}{2n^2}\mu$ until $n$
decreasing to the lowest limit $n=1$. The Lamb shift should be
viewed as a further modification on the mass of an off-mass-shell
electron due to radiative correction.

 Notice that the parameter $Q^2$ in the vertex function,
 Eq.(\ref{6-14}), means the square of (three-dimensional) momentum
 transfer when a free electron is on its mass-shell and collides
 with some other particle as discussed in Ref.\cite{8:b}. By contrast, now
 $Q^2$ exhibits itself as an effective potential of
 $\delta$-function type exerted by the nucleus to the bound (and
 so off-mass-shell) electron as shown by Eq.(\ref{7-5}). To bind an electron to a nucleus is a
 nonperturbative effect. Hence we can understand why the discrepancy between $\zeta^{<S>}$
 (calculated by perturbative QED at one-loop order) and $\zeta^{<V>}$ (evaluated via nonperturbative
 Virial theorem) is so large. Fortunately, they lead to discrepancies in the calculated values of Lamb shift being not so large as
 shown in Section VIII. When $\zeta^{<V>}$ or $\zeta^{<S+V>}$ (or $\zeta^{<SV>}$) is substituted into the Eq.(\ref{8-4})
 which is derived from perturbative ($L=1$) theory, we should always be aware of some theoretical inconsistency in such a
 semi-empirical treatment. But as a whole, we believe that the concept of Lamb shift as an off-mass-shell effect in covariant QED
 is basically correct.

 (c) For a free on-mass-shell electron, its charge square $e_R^2$
 will increase with the increase of $Q^2$ as shown by
 Eq.(\ref{5-10}) (with $\mu_3=m_e,\ q^2=-Q^2$) and was calculated
 in detail in \cite{8:b}, coinciding with the experimental data.
 Note that, however, the Ward identity $Z_1=Z_2$ had been
 used. An interesting question arises for a bound electron: as its
 $e_R^2$ is not a function of $Q^2$, will $e_R^2$ change with the
 variation of the quantum number $n$? To answer this question, let us
 put Ward identity aside for a while and write down the
 renormalized $\alpha_R=\frac{e_R^2}{4\pi}$ as

 \beq
\alpha_R=\frac{Z_2^2}{Z_1^2}Z_3\alpha\longrightarrow\frac{Z_2^2}{Z_1^2}\alpha
\label{9-1} \eeq

Let us work in the CMCS, so $Z_2=\frac{1}{1-B}$ and $B$ is shown
in Eq.(\ref{4-8}) but with $\mu$ replaced by $m_e=m$. Similarly,
$Z_1$ is given by the first part of Eq.(\ref{6-14}) with
$\mu\longrightarrow m$:

\beq Z_1\simeq
1+\frac{\alpha}{4\pi}[\frac{11}{2}-\ln\frac{m^2}{\mu_1^2}-3\xi+4(1+\xi)\ln\xi]
\label{9-2} \eeq

where the off-mass-shell parameter $\xi$ in CMCS is defined by

\beq
p^2=m^2(1-\xi)=m^2(1-\eta-\zeta')=m^2(1-\eta)-m^2\zeta'=\mu^2-m^2\zeta'=\mu^2(1-\zeta)
\label{9-3} \eeq

with

\beq m^2(1-\eta)=\mu^2, \ \ \eta=1-\frac{\mu^2}{m^2},\ \
\zeta'=\frac{\mu^2}{m^2}\zeta \label{9-4} \eeq

and $\zeta$ is exactly that in Eq.(\ref{7-1}). If we ignore the
dependence of $(1-B)$ on $\zeta$, Eq.(\ref{9-1}) would give
($\zeta\ll1$):

\beq
\alpha_R=\alpha[1+\frac{2\alpha}{\pi}\ln(1+\frac{\zeta}{\eta})]
\label{9-5} \eeq

after renormalization by adjusting the arbitrary constant $\mu_1$
so that

\beq \alpha_R|_{\zeta\rightarrow0}=\alpha \label{9-6} \eeq

which connects to the Thomson limit
$\alpha_R|_{Q\rightarrow0}=\alpha$ for a free electron
continuously but not smoothly. Then for two lowest bound
states with $n=1$ and $n=2$, we would have (in $<V>$ scheme):

\beq \alpha_R|_{n=1}\simeq\alpha(1.000433832),\ \
\alpha_R|_{n=2}\simeq\alpha(1.0001123)\label{9-7} \eeq

This would modify the Bohr energy level in hydrogenlike atom A to

\beq \tilde{E}_A^{Bohr}(n)=-\frac{Z^2\alpha_n^2}{2n^2}\mu_A
\label{9-8} \eeq

and make an extra contribution to the isotope-shift as

\beq \Delta \tilde{E}_{D-H}^{Bohr}(2S-1S)\simeq726\ \ MHz
\label{9-9} \eeq

which is definitely excluded by the experiment. Hence the above
consideration from Eq.(\ref{9-1}) till Eq.(\ref{9-9}) is wrong. We
learn concretely once again that the Ward identity $Z_1=Z_2$ is valid not
only for an electron on its mass-shell, but also for
off-mass-shell case. Thus we use the same value of $\alpha$
throughout the whole calculation.

(d) In Ref.\cite{8:b}, using our RRM and new renormalization group equation (RGE) for QCD derived from it and keeping all masses of 6 quarks ($m_c=1.031\; GeV$, $m_b=4.326\; GeV$,
$m_t=175\; GeV$, $m_s=200 \;MeV$, $m_u=8\; MeV$, $m_d=10\; MeV$), we calculated the strong coupling constant $\alpha_{s_i}(Q)$ for $i=u,d,s,c,b$ respectively.
Their running curves (starting from the common renormalization point $\alpha_s(M_Z)=0.118$) follow the trend of experimental data
(as shown on p.158 of \cite{39}) quite well but separate at the low $Q$ region. Interesting enough, each of them rises to a maximum
$\alpha_{s_i}^{max}$ and then suddenly drops to zero at $Q=0$ corresponding to a threshold energy scale $E_i^{th}$ which could
be explained as the excitation energy scale for breaking the quark pair. For example, we find $E_b^{th}=1.13\; GeV$ which is
just the hadronization energy scale of Upsilon $\Upsilon(b\bar{b})$ against its dissociation into two bosons. Experimentally,
$M(\Upsilon(4s))-M(\Upsilon)=1.12\;GeV$ and $\Upsilon(4s)\to B^+B^-$ or $B^0\bar{B}^0$. similarly, $E_c^{th}= 0.398\;GeV$
while $M(\psi(3770))-M(\psi(3097))=673 \;MeV$ and $\psi(3770)\to D^+D^-$ or $D^0\bar{D}^0$. it seems that
$E_s^{th}\sim 90\,MeV$ and $E_{u,d}^{th}\sim 0.4\,MeV$ are not so reliable but still reasonable.

Actually, our calculation on QCD is backed by that on QED. In [8]b, using our RRM and improved RGE, keeping all contributions from 9
charged leptons and quarks we were able to calculate the running fine-structure constant $\alpha_R(Q)$ from the renormalization point
$\alpha_R(Q)|_{Q=0}=\alpha=(137.036)^{-1}$ until it coincides with the experimental value of $\alpha_{exp}(M_Z)=(128.89)^{-1}$.
We fitted quark's masses as mentioned above and found no further room left for extra charged elementary particles (say, of 4th
generation).

(e) In 1989, we had estimated the upper and lower bounds on Higgs mass $M_H$ by using a nonperturbative approach in QFT --- the Gaussian effective
potential (GEP) method, yielding \cite{40}:
\begin{equation}\label{}
76\; GeV <M_H<170\; GeV
\end{equation}
Like many other authors, we were bothered a lot by divergences. After a deeper study on the $\lambda\phi^4$ model by using our new RRM
\cite{8:a}, we restudied this problem by combination GEP with RRM, yielding\cite{24}:
\begin{equation}\label{higgs}
M_H=138\; GeV
\end{equation}
This is not a upper or lower bound but a prediction based on the input of experimental data:
\begin{equation}\label{}\begin{array}{l}
M_W=80.359 \;GeV,\;M_Z=91.1884 \;GeV, \alpha^{-1}=\dfrac{4\pi}{g^2\sin^2\theta_W}=128.89,\;\\
\sin^2\theta_W=\dfrac{{g'}^2}{g^2+{g'}^2}=0.2317
 \end{array}\end{equation}
where $\theta_W$ is the weak mixing (Weinberg) angle. Because of getting rid of all divergences, our calculation is well under
control at every step. As now the search for Higgs particle becomes so urgent experimentally but the theoretical estimation about
its mass still remains uncertain\cite{39}, we think our approach with the prediction (\ref{higgs}) deserves to be reconsidered.

(f) Moreover, our RRM can be used in $D+1$ space-time without
limitation on the space dimension $D$. A detailed analysis of
sinh(sine)-Gordon models with $D=1,2$ and $3$ (also using GEPM) is
given by Ref.\cite{25}. Another example is again the Lamb shift but
calculated by QED in noncovariant form and by using RRM similar to
that in this paper, see Appendix (\cite{26}, see also the Appendix 9A in
\cite{15}). The theoretical value (A.20) seems better than (\ref{8-20}), showing that for
dealing with the Lamb shift, the noncovariant theory may be more suitable than the
covariant one at least in the lowest order.

(g) Previously, the theories for Lamb shift or generally for
calculating energy levels in hydrogenlike atoms are rather complicated
as reviewed in refs \cite{1,2} and \cite{27}, some of them have been discussed
in the Appendix of this paper. For further clarity, let us try to
summarize the main obstacles, or challenges in four points:\\
(1) Different masses of nuclei must be taken into account;\\
(2) Relativistic effects of the electron (not nucleus) are important;\\
(3) In calculating radiative corrections, the divergence becomes
severer and severer with the increase of loop number;\\
(4) Since nuclei's properties are different from one atom to another,
to treat each atom as a two-body system individually would be a
daunting task, it couldn't be rigorous eventually too. This can be
clearly seen from the recent work by Marsch \cite{34}.

   Facing these challenges and learning from lessons and experiences
of previous authors, we see that the clue point is to replace the
electron mass $m$ by reduced mass $\mu$ and work in the noninertial frame
(RMCS) throughout the entire calculation. As is well known, this can
be handled in nonrelativistic QM by a
mathematical trick but is impossible in relativistic case. So what we
need is a new understanding on the essence of special relativity
--- the invariance (of theory) under the (newly defined) space-time
inversion in one inertial frame. Then we are able to claim the same
invariance in the RMCS with $\mu$ replacing $m$
for establishing the RDE, ignoring a small centripetal acceleration of the nucleus in $CMCS$ (see page note after Eq.(\ref{3-22})).
The approximation in $RDE$ is some price paid for the much
bigger gain---improving the original Dirac equation (unable to treat
different nuclei) and avoiding the confusion in QED calculation
because of the entanglement of two frames: CMCS with RMCS (
{\it i.e.}, the radiative corrections are entangled with the recoil effect
as we can see from previous literatures). In some senses, we jump over
obstacles (1) ,(2) and (4)at the least labor cost (by constructing
RDE). In the meantime, we hope RDE would help to ease the
difficulty in point (3).
And it's a great pleasure to see that the essential correctness of our
understanding has been validated by Marsch's work as well as
puzzles raised in his paper \cite{34}. Please see also Ref \cite{41}.

   As to challenge (3), only after we puzzled over the "divergence"
for decades, could we suddenly realize that we misread its implication
as a "large number".
Rather, it means the "uncertainty". Let us look at the calculation in
section IV again. Previously, many authors treated the divergent
integral I in Eq.(\ref{4-5}) by different tricks of regularization ,
arriving at Eq.(\ref{4-11}). Because both A and B are divergent, it was
thought that the original mass ($\mu$ here) does receive some radiative
corrections (via the self-energy diagrams in Fig 2(a) and (b)) and
becomes a "renormalized" mass ($\mu_R$ here). The latter should be the
observed mass in experiment or physical mass (of electron). So the
original mass was called as the "bare mass", which was written into
the Lagrangian density as an input parameter of QFT. Then in
constructing Feynman diagrams of certain perturbative calculation, one
needs to further introduce some (divergent) "counter terms" for
cancelling the divergence stemming from the bare mass. Based on that understanding, the renormalization
factor for wavefunction, $Z_2$ in Eq.(\ref{4-10}), would be a divergent quantity
too (in sharp contrast to here Eq.(\ref{4-14}) being a fixed
and finite number). Previously, In Eq.(\ref{5-10}), while the $e_R$ on the left
handed side is the observed charge which should be finite, the $e$ on
the right handed side was regarded as a "bare charge" which, together
with the $Z_3$, was a divergent quantity. (see Fig.~7.8 in \cite{23}. By contrast, here both $Z_3$ and
$e$ are finite. Actually, here $e$ is defined as the physical charge
observed at the Thomson limit in experiment).

  In our opinion, the reason why we encountered so many  superfluous troubles
in the past is because we overlooked what Bethe said in 1947\cite{29}.
Please read his words quoted after Eq. (A.2) in the Appendix. Let us
explain via our Eq.(\ref{4-1}).
The (reduced) mass ($\mu$) already contains some contributions from
self-energy diagrams like Fig. 2(a) and (b). When we evaluate the
(divergent) integral, Eq.(\ref{4-5}), trying to find the radiative
corrections on the electron, the latter is bound to confuse with that
already contained in the mass. In other words, the dividing line
between them is blurred inevitably. The emergence of explicit
divergence is essentially a warning that the effect you want to
evaluate has been entangled with the mass itself, rendering both of
them uncertain. Hence the aim of so-called renormalization is nothing
but to redraw the dividing line between them such that the values of
mass (reconfirmed by the experiment) and the new effect ({\it e.g.}, the
mass increment when the electron is moving off-mass-shell, Eq.(\ref{4-16}))
can be clearly separated.
  In short, what we have been learning in the past decade is:
At the level of QM, in the Hamiltonian like Eq. (A.1), the parameters
$m$ and $e$ can be regarded as well-defined. But they are not so at the
level of QFT. Once the calculation is made beyond the tree level,
{\it i.e.}, with loop number $L\geq1$, the divergence occurs and the meaning of
parameters becomes obscure immediately.

We need to reconfirm all parameters contained in the Lagrangian
density before they can be linked with experiments. In this sense, a
model of QFT is at most an "effective field theory". According to the
above point of view, we believe that our RRM just provides a natural
way to carry out these processes of reconfirmation \cite{8:a}, getting rid
of divergences and ambiguities. Please see also Ref \cite{42}.

(h) Last, but not least, during the learning and teaching of graduate courses
on QFT for decades, we have been sharing the joy and puzzle with
our students all the time. We hope that the presentation of this
paper could be useful as a teaching reference to render the QFT
course more understandable, interesting and attractive.

\section*{Acknowledgements}

We thank S.Q. Chen, S.S. Feng, R.T. Fu,
P.T. Leung, W.F. Lu, X.T. Song, F. Wang, H.B. Wang, J. Yan, G.H. Yang and
J.F. Yang for close collaboration and helpful discussions.
We are also indebted to referees whose comments provided us opportunities to
improve our manuscript.

\section*{Appendix: Comparison Between Noncovariant and Covariant Theories for Lamb Shift}
1. To our knowledge, the precision theory for Lamb shift was based on a combination of noncovariant
(nonrelativistic or old-fashioned) QED with covariant (or relativistic) QED as discussed in Ref.\cite{27}.
As explained clearly by Sakurai in Ref.\cite{21}, in perturbative QFT of noncovariant form, all virtual
particles are "on-mass-shell". Here we wish to emphasize that a rigorous reconfirmation procedure of mass
parameter was often overlooked in previous literatures.
The theory for hydrogenlike atom begins with a Hamiltonian:
\begin{equation*}\label{A1}
    H_0=\frac{1}{2m}{\mathbf p}^2+\frac{1}{2m_N}{\mathbf p}^2-\frac{Z\alpha}{r}\eqno{(A.1)}
\end{equation*}
(${\mathbf p}=-i\nabla$, see Eq.(34) in \cite{27}). As Bethe \cite{29} first pointed out that the effect of
electron's interaction with the vector potential $\mathbf A$ of radiation field (see \cite{21},\cite{15})
\begin{equation*}\label{A2}
    H_{int}^{(1)}=\frac{e}{mc}{\mathbf A}\cdot{\mathbf p}\eqno{(A.2)}
\end{equation*}
should properly be regarded as already included in the observed mass $m_{obs}$ of the electron, which is denoted
by $m$ in (A1). However, once a concrete calculation is made with (A2) being taken into account, the
divergence emerges immediately. What does it mean?
Mathematicians teach us that there are three implications for a divergence:\\
(a)It is a dimensionless number; (b)It is a large number; (c)It is uncertain. While we physicists often emphasized
the point (b), we didn't pay enough attention to the points (a) and (c). We often talked about a quadratically
(or linearly) divergent integral without noticing that it has a dimension (say, mass dimension) and thus
meaningless in mathematics unless a mass parameter (say, $m$) in the integral is already fixed as a mass "unit"
so that the integral can be divided by $m^2$ (or $m$) to become dimensionless.
Alternatively, a logarithmically divergent integral is dimensionless and thus unaffected by the choice of unit
[like Eq.(\ref{4-5}), see also Eq.(A6) below], it just implies an uncertainty waiting to be fixed.
The implication of uncertainty of a divergence will never vanish even after we introduced a cutoff by hand to
curb it. For example, in a pioneering paper to explain the Lamb shift, Welton (\cite{30}, see section 9.6B in
Ref.\cite{15}) encountered an integral $I=\int_{\omega_{min}}^{\omega_{max}}\frac{d\omega}{\omega}$ with $\omega$
being the (angular) frequency of virtual photon (vacuum fluctuation). He simply set
$\omega_{min}\sim mZ\alpha =Z/a$, ($a$ is Bohr radius) and $\omega_{max}\sim m$ so that
$I\simeq \ln(1/Z\alpha)=4.92$ (for $Z=1$) which leads to an estimation of Lamb shift
$L_H^{theor.}(2S_{1/2}-2P_{1/2})\simeq 668\ MHz$. If instead of Bohr radius, the lower cutoff is provided by the
electron binding energy, one should get $I\simeq \ln(Z\alpha)^{-2}$ and $L_H^{theor.}\simeq 1336\ MHz$.
(see Eq.(30) in \cite{27}).
The above arbitrariness just reflects what essential in a divergent integral is not its large magnitude
($\ln(Z\alpha)^{-1}$ is merely of the order of 10) but its uncertainty. So what important in handling the integral
is not to curb (or to hide) its divergence but let the divergence exhibits itself as some arbitrary constants
explicitly (as shown in section IV-VI). We will show later how to do this way for noncovariant QED.
\\
2. While Eqs.(A1) and (A2) only describe a spinless particle, the electron has spin which endows it
with the relativistic nature as shown by Eqs.(\ref{3-21})-(\ref{3-25}). For two-particle system,
based on Bethe-Salpeter
equation, an effective Dirac equation (EDE) was derived as shown by Eq.(23) in \cite{27}. When the electromagnetic
field interaction is taken into account, the Breit potential $V_{Br}$ was derived as shown by Eq.{(35)}
in \cite{27}.
Then the total Breit Hamiltonian reads (Eq.(36) in \cite{27}):
\begin{equation*}\label{A3}
    H_{Br}=H_0+V_{Br}\eqno{(A.3)}
\end{equation*}
However, the electron mass $m'$ (in our notation here) appeared in EDE or $V_{Br}$ should be that in the Dirac
equation, also that in the definition of reduced mass $\mu=\frac{m'm_N}{m'+m_N}$, eventually $m'$ could be
identified with the observed mass $m_{obs}$, which is not equal to the $m$ in Eq.(A1). This is because besides
(A2) there is an extra interaction due to electron spin with the radiation field:
\begin{equation*}\label{A4}
    H_{int}^{(2)}=\frac{ge\hbar}{4\mu c}{\mathbf \sigma}\cdot\nabla\times {\mathbf A}\eqno{(A.4)}
\end{equation*}
($g=2\times 1.0011596522$ is the gyromagnetic ratio of electron, see Eq.(9A.15) in \cite{15}). The
difference between
 $m$ and $m'$ will be calculated in (A16) below. It turns out to be
of the order of $\alpha m$ and cannot be ignored at the level of QED, especially for the explanation of
Lamb shift.
We guess this must be one of the reasons why all calculations based on Eq.(A3) became so complicated.
\\
3. In noncovariant theory, the leading contribution to the Lamb shift comes from the one-photon electron
self-energy.
The nomenclature here is different from that in the covariant theory. Roughly speaking, so-called electron
self-energy often corresponds to the vertex function in covariant theory (Fig.2(d) in this paper)
or to Figs.8 and 11
in Ref.\cite{27} and its evaluations have extended over 50 years \cite{31}. More precisely, it is identified with
the radiative insertions in the electron line and the Dirac form factor contribution. Further contributions from
the Pauli form factor and the vacuum polarization \cite{27} will add to a theoretical value of classic Lamb shift
being $1050.559\ MHz$.
If taking more high-order corrections into account, the theoretical value coincides with the experimental value
$1057.845\ MHz$ rather accurately (see Table 20 in \cite{27}).
However, the above calculation looks quite complicated due to two reasons: (a) The difficulty of dealing with
two masses in two coordinate systems, the electron mass $m$ and the reduced mass $\mu$; (b) The introduction of
an auxiliary parameter $\sigma\ [m(Z\alpha)\gg \sigma\gg m(Z\alpha)^2]$ to separate the radiative photon
integration
region into two parts. In the low momentum region, the Bethe Logarithm \cite{32} in noncovariant form makes the
main contribution. In the high momentum region, the evaluation is resorting to some relativistic covariant form
\cite{22}. Then two expressions are matched together to get the correct result.
It seems to us that the matching trick used is doubtful because both ultraviolet and infrared divergences were
ambiguously handled by some cutoff which missed the main point of renormalization---to reconfirm the mass parameter
in the presence of radiative corrections as shown in section IV (covariant form) or below.
\\
4. A simple calculation for Lamb shift in noncovariant form was proposed in Ref.\cite{26} (see also Appendix
 9A of Ref.\cite{15}). Consider the self-energy diagram of an electron with reduced mass $\mu$ and
 (three-dimensional)
 momentum $\mathbf p$ in the RMCS of a hydrogenlike atom. Similar to Fig.2(a), but also different in the virtual
 state, now a photon has energy $\omega_k=k=|{\mathbf k}|$ while the electron has momentum
 ${\mathbf q}={\mathbf p}-{\mathbf k}$ and energy $\varepsilon_q=\frac{1}{2\mu}q^2$.
 The electron in plane-wave state $|{\mathbf p}>$ has two interactions with the radiative field at each vertex as
 shown by (A2) ($m\rightarrow \mu$) and (A4), acquiring an increase in energy respectively (see FIG. 3):
\begin{equation*}\label{A5}
    \Delta E_p^{(j)}=\sum_i\frac{|<i|H_{int}^{(j)}|{\mathbf p}>|^2}{\varepsilon_p-\varepsilon_i},
\quad (j=1,2)\eqno{(A.5)}
\end{equation*}
Here $\varepsilon_i=\varepsilon_q+\omega_k$ is the energy of the intermediate virtual state $|i>$.
Simple evaluation leads to
\begin{equation*}\label{A6}
    \Delta E_p^{(1)}=-\frac{\alpha p}{\pi\mu}\int_{-1}^1d\eta(1-\eta^2)I,
\quad I=\int_0^{\infty}\frac{dk}{k+\xi}\eqno{(A.6)}
\end{equation*}
where $\eta=\cos\theta$ with $\theta$ being the angle between $\mathbf k$ and $\mathbf p$, $\xi=2(\mu-p\eta)$.
Like Eq.(\ref{4-5}), we take partial derivative of the divergent integral $I$ with respect to $\xi$ (then the
integration of $k$) and integrate back to $I$ again, yielding:
$$
  \Delta E_p^{(1)} = b_1^{(1)}p^2+b_2^{(1)}p^4+\cdots \eqno{(A.7)}
$$

  $$b_1^{(1)} = \frac{\alpha}{\pi\mu}(\frac{4}{3}\ln2+\frac{4}{3}\ln\mu-\frac{4}{3}C_1) \eqno{(A.8)}$$

  $$b_2^{(1)} = \frac{\alpha}{\pi\mu^3}(-\frac{2}{15})\eqno{(A.9)}$$
Note that the term $b_1^{(1)}p^2$ will combine with the kinetic energy $\frac{1}{2\mu}p^2$ of a ("spinless")
electron, they are indistinguishable. The appearance of an arbitrary constant $C_1$ precisely reflects the fact
that we cannot find the reduced mass via the valuation of $\Delta E_p^{(1)}$ in perturbation theory. So we must
choose $b_1^{(1)}=0$ to reconfirm the value of $\mu$ (which is still not the final observed mass, see below).
Similar evaluation on $H_{int}^{(2)}$ (of the real electron with spin $1/2$) which would induce the spin flip
process between states $|{\mathbf p},\pm\frac{1}{2}>$ and $|{\mathbf q},\pm\frac{1}{2}>$, leads to
$$
  \Delta E_p^{(2)} =\frac{1}{2}\sum_{i,s_z=\pm1/2}\frac{|<i|H_{int}^{(2)}|{\mathbf p},s_z>|^2}
  {\varepsilon_p-\varepsilon_i}=-\frac{\alpha g^2}{8\pi\mu}\int_{-1}^1d\eta J
$$

$$  J =\int_0^{\infty}\frac{k^2dk}{k+\xi}\eqno{(A.10)}$$
Being a quadratically divergent integral, $J$ needs partial derivative of third order with respect to $\xi$,
yielding:
$$
 \Delta E_p^{(2)} =b_0^{(2)}+b_1^{(2)}p^2+b_2^{(2)}p^4+\cdots \eqno{(A.11)} $$

 $$
 b_0^{(2)} = \frac{g^2}{4}\frac{\alpha\mu}{\pi}[4(\ln2+\ln\mu)-4C_2-\frac{2C_3}{\mu}-\frac{C_4}{\mu^2}]\eqno{(A.12)}
 $$

 $$
  b_1^{(2)} =\frac{g^2}{4}\frac{\alpha}{\pi\mu}(\frac{4}{3}\ln2+2+\frac{4}{3}\ln\mu-\frac{4}{3}C_2) \eqno{(A.13)}
  $$

 $$ b_2^{(2)} = \frac{g^2}{4}\frac{\alpha}{\pi\mu^3}(-\frac{1}{15})\eqno{(A.14)}$$
Let's manage to fix three arbitrary constants $C_2,C_3$ and $C_4$. First, the term $b_1^{(2)}p^2$ should be
combined
with $\frac{1}{2\mu}p^2$ term. Since $\mu$ is already fixed, further modification on $\mu$ due to electron spin
should be finite and fixed. So the only possible choice of $C_2$ is to cancel $\ln\mu$ which is ambiguous in
dimension: $C_2=\ln\mu$, yielding
\begin{equation*}\label{A15}
    b_1^{(2)}=\frac{\beta}{2\mu},\quad \beta=\frac{g^2\alpha}{2\pi}(\frac{4}{3}\ln2+2)\eqno{(A.15)}
\end{equation*}
Then the dimensional constants $C_3$ and $C_4$ must be chosen such that $b_0^{(2)}=0$, implying that the starting
point of this theory is the nonrelativistic Hamiltonian $H_0$ in Eq.(A1) without rest energy term while both
masses of the nucleus and the electron (with spin) are fixed by experiments.
Hence now $\mu$ acquires a modification via $b_1^{(2)}p^2$ term and becomes an observable one:
\begin{equation*}\label{A16}
   \mu\longrightarrow \mu_{obs}=\frac{\mu}{1+\beta}\eqno{(A.16)}
\end{equation*}
However, we have to consider the relativistic energy of electron shown in Eq.(\ref{3-10}), where the term
$(-\frac{1}{8\mu^3}p^4)$ goes beyond Eq.(A1). Yet the modification of $\mu$ shown as (A16) does induce
a corresponding change $-\frac{1}{8}(\frac{1}{\mu_{obs}^3}-\frac{1}{\mu^3})p^4$, which should be regarded as
an invisible "background" and subtracted from the $p^4$ term induced by radiative corrections. (The relativistic
correction is brought in via the RDE as discussed in section VIII). As a whole, the combination of contributions
from $H_{int}^{(1)}$ and $H_{int}^{(2)}$ leads to
\begin{equation*}\label{A17}
   b_1= b_1^{(1)}+ b_1^{(2)}= b_1^{(2)}\eqno{(A.17)}
\end{equation*}
and a "renormalized" $b_2$:
\begin{equation*}\label{A18}
   b_2^R=b_2^{(1)}+b_2^{(2)}+\frac{1}{8\mu^3}(3\beta+3\beta^2+\beta^3)
   \simeq\frac{\alpha}{\pi\mu_{obs}^3}(1.99808)\eqno{(A.18)}
\end{equation*}
Here we only keep the lowest approximation at the last step.
Hence the electron self-energy-diagram contributes a radiative correction to the energy level of the
stationary state $|Z,n,l>$ in a hydrogenlike atom:
\begin{equation*}\label{A19}
   \Delta E^{rad}(Z,n,l)=\langle Z,n,l|b_2^Rp^4|Z,n,l\rangle=[\frac{8n}{2l+1}-3]\frac{b_2^RZ^4\alpha^4}{n^4}\mu^4_{obs}\eqno{(A.19)}
\end{equation*}
This form, together with contributions from the vacuum polarization and nuclear size effect, gives a theoretical
value for classic Lamb shift:
\begin{equation*}\label{A20}
   L_H^{theor.}(2S_{1/2}-2P_{1/2})\approx 1056.52\ MHz\eqno{(A.20)}
\end{equation*}
which is smaller than the experimental value by $0.13\%$.
Despite its approximation involved, the above method clearly shows that so-called renormalization is nothing but a
reconfirmation process of mass. We must reconfirm the mass before it could be modified via radiative corrections.
Either "skipping over the first step" or "combining two steps into one" is not allowed.
\\
5. In noncovariant theory, the (three-dimensional) momentum $\mathbf p$ is combined with the reduced mass $\mu$
to form a kinetic energy term $\frac{1}{2\mu}{\mathbf p}^2$ on the mass shell. Once the energy is modified
whereas $\mathbf p$ is conserved at the vertex, $\mu$ is bound to be modified. On the other hand, in covariant
theory, the electron energy turns to a component of four-dimensional momentum $p$ and the latter is conserved
at the vertex. So the (reduced) mass  $\mu$  cannot be modified on the mass shell $(p^2=\mu^2)$. Therefore,
the renormalization as some reconfirmation has different meaning in covariant theory versus that in
noncovariant theory. We guess this is why the matching procedure of these two formalisms into one theory
for Lamb shift proves so difficult.
\\
6. Every theory in physics is not only a discovery of natural law, but also an invention of human being \cite{33}.
Hence the comparison among various theories, in many cases, is not about a problem of being right or wrong.
Rather, it's about a choice of simplicity, harmony (self-consistency) and beauty. Only time can tell.


\begin{figure*}[!h]
\centerline{\includegraphics[scale=0.8]{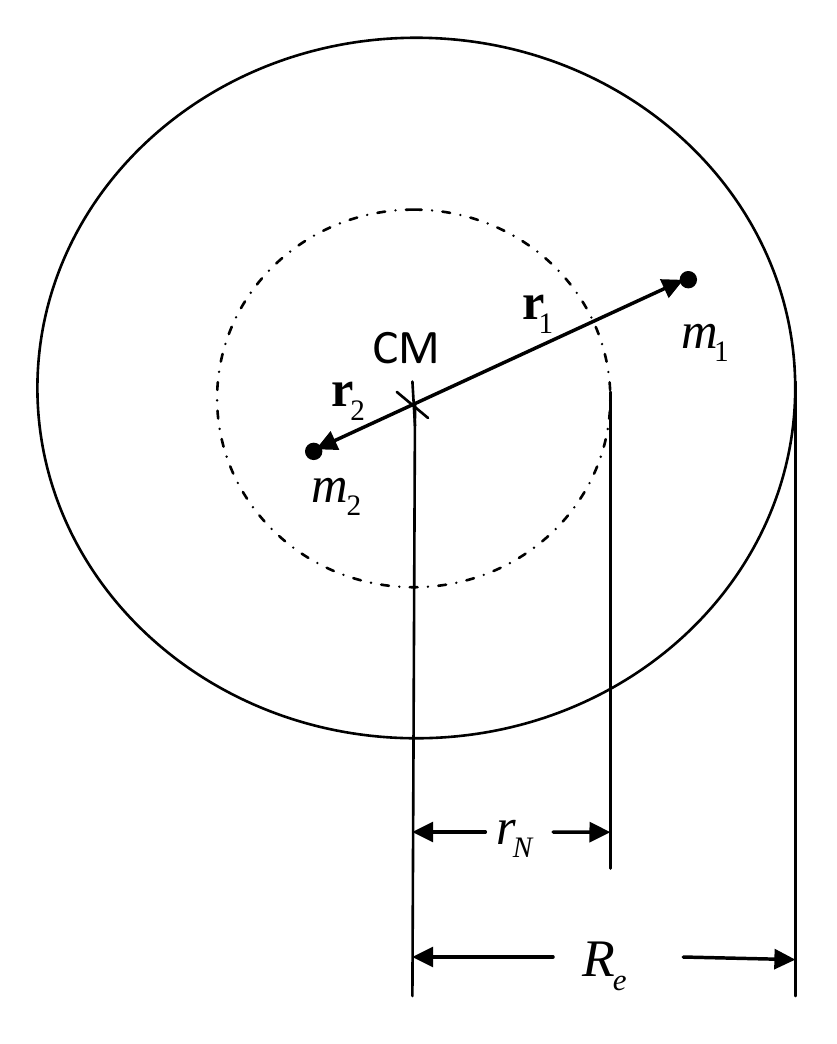}}
\vskip 0.5cm
\caption{
A hydrogenlike atom in quantum mechanical description. The nucleus with mass $m_2$ occupies a small sphere with radius $r_N$
(greatly exaggerated in the diagram) while the electron with mass $m_1$ spreads over a larger sphere with radius $R_e$
(\ie atomic radius). Their common center is the atom's center of mass (CM). The wavefunction $\psi({\bf r})e^{-iEt}$ with
${\bf r}={\bf r}_1-{\bf r}_2$ shows the electron's amplitude under a "fictitious measurement" \cite{15}, during which the
electron and nucleus shrink into two "fictitious point particles " located at ${\bf r}_1$ and ${\bf r}_2$ simultaneously.
The Coulomb potential $V(r)=-\frac{Ze^2}{r}$ between them is a static one. The probability
to find the electron at ${\bf r}$ is $|\psi({\bf r})|^2$ while that to find its momentum being ${\bf p}$ is $|\phi({\bf p})|^2$ with
$\phi({\bf p})$ being the Fourier transform of $\psi({\bf r})$.
}
\end{figure*}

\begin{figure*}[!h]
\centerline{\includegraphics[scale=0.8]{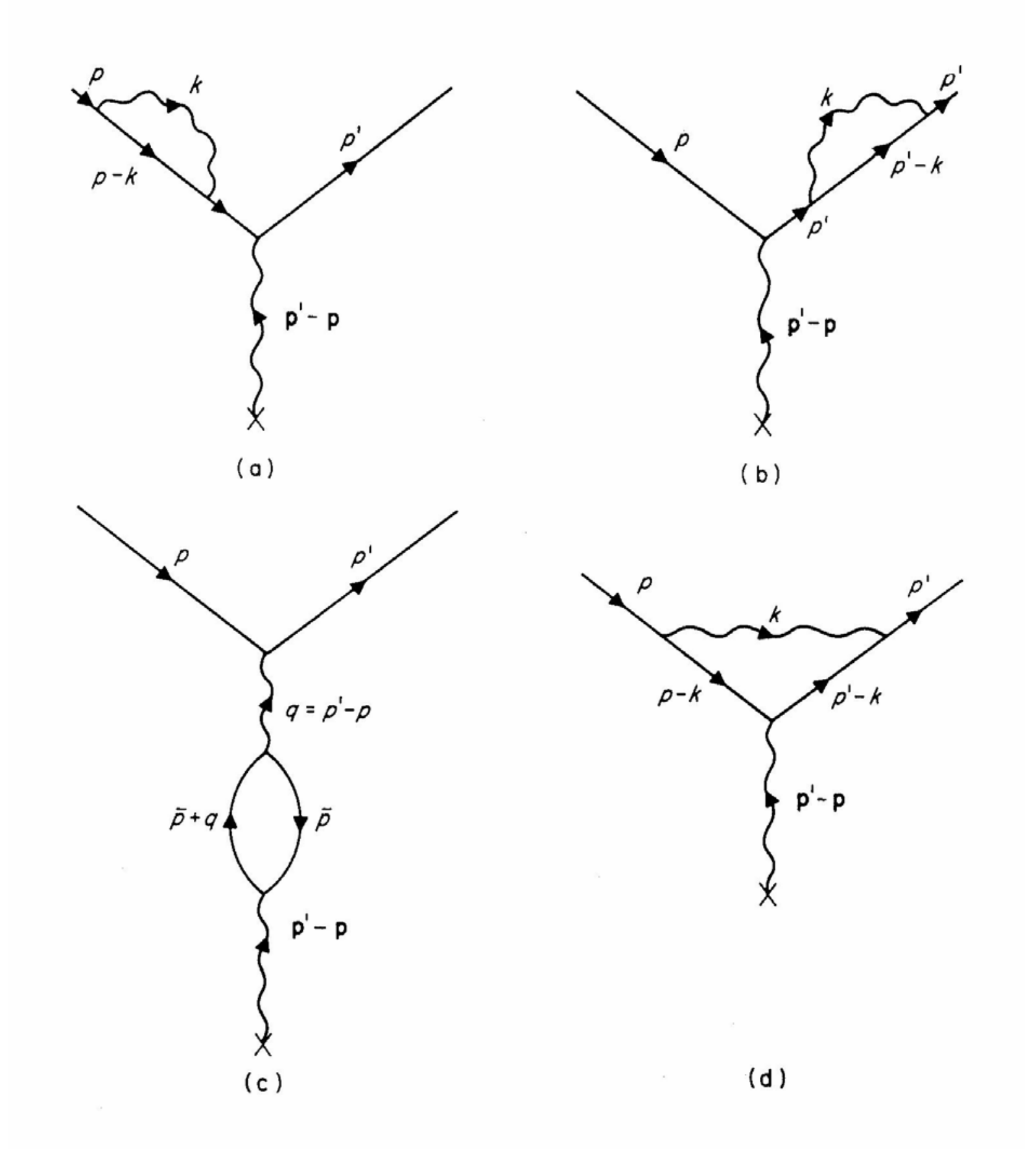}}
\vskip 0.5cm
\caption{Four Feynman diagrams at one-loop level (in covariant form). (a) and (b) are self-energy diagrams of the electron.
(c) is vacuum polarization. (d) is vertex function. Solid lines and wavy lines refer to electron and photon
respectively, while X denotes the nucleus. Here $p,q$ and $k$ are
four-dimensional momenta.}
\end{figure*}

\begin{figure*}[!h]
\centerline{\includegraphics[scale=1.3]{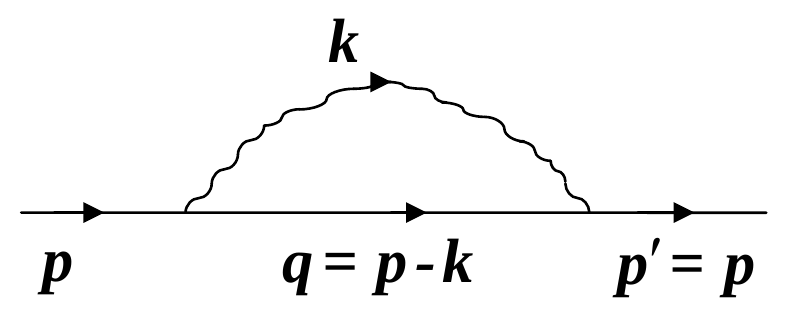}}
\vskip 0.5cm
\caption{
The electron self-energy (radiative correction) diagram at one-loop level of perturbative QCD in noncovariant form.
$H_{int}^{(1)}$ (A.2) or $H_{int}^{(2)}$ (A.4) is inserted into two vertices. Here $\bf p,q$ and $\bf k$ are three
dimensional momenta.
}
\end{figure*}

\end{document}